\documentclass[vecphys]{svmult}

\usepackage{makeidx}         
\usepackage{graphicx}        
\usepackage{multicol}        
\usepackage[bottom]{footmisc}

\makeindex             

\begin{document}

\title*{Markov-Chain Monte Carlo Methods for Simulations of 
Biomolecules}
\author{Bernd A. Berg\inst{1,2}}
\institute{School of Computational Science (SCS), Florida State 
University, Tallahassee, FL 32306-4130, USA \texttt{berg@scs.fsu.edu}
\and Department of Physics, Florida State University,
Tallahassee, FL 32306-4350, USA \texttt{berg@hep.fsu.edu}}
%
%
\maketitle

The computer revolution has been driven by a sustained increase of 
computational speed of approximately one order of magnitude (a factor 
of ten) every five years since about 1950. In natural sciences this 
has led to a continuous increase of the importance of computer 
simulations. Major enabling techniques are Markov Chain Monte 
Carlo (MCMC) and Molecular Dynamics (MD) simulations. 

This article deals with the MCMC approach. First basic simulation 
techniques, as well as methods for their statistical analysis are 
reviewed.  Afterwards the focus is on generalized ensembles and 
biased updating, two advanced techniques, which are of relevance 
for simulations of biomolecules, or are expected to become relevant 
with that respect. In particular we consider the multicanonical 
ensemble and the replica exchange method (also known as parallel 
tempering or method of multiple Markov chains).

\section{Introduction}\label{sec_Introduction}


Markov chain Monte Carlo (MCMC) calculations started in earnest with 
the 1953 paper by Nicholas Metropolis, Arianna Rosenbluth, Marshall
Rosenbluth, Augusta Teller and Edward Teller~\cite{Me53}. Since then 
MCMC simulations have become an indispensable tool with applications
in many branches of science. Some of those are reviewed in the proceedings
\cite{Gu04} of the 2003 Los Alamos conference, which celebrated the 50th 
birthday of Metropolis simulations. 

The purpose of this article is to give a concise overview ranging from 
statistical preliminaries and plain Monte Carlo (MC) calculations over 
basic techniques for MCMC simulations to advanced methods, which are 
indispensable when it comes to the simulations of complex systems with 
frustrated interactions. For such systems rugged free energy landscapes 
are typical. Here our focus is on biomolecules such as peptides (small 
proteins) in an all-atom approach, defined by a model energy function. 
At a given temperature this energy function determines in principle the 
Gibbs ensemble of the molecule. In practice equilibrium is sometimes 
hard to reach in MCMC simulations of the canonical ensemble. 
Considerable improvements can be made by using generalized ensembles 
and biased sampling.

The first part of this article gives a treatment of the MCMC 
fundamentals that is largely based on the author's book \cite{BBook} 
on the subject, which gives many more details and contains extensive 
additional material. The book comes with computer code that can 
be downloaded from the web. The solutions of numerous numerical 
assignments, are reproducible by compiling and running the 
corresponding computer programs. Informations and a link to the 
computer code are found on the web at 
\texttt{http://www.scs.fsu.edu/\~\ $\!\!\!$berg}.

The second part of this article builds on original literature 
of generalized ensemble methods \cite{ToVa77,SwWa86,Ge91,BeNe91,
BeNe92,Be92,LyMa92,MaPa92,HuNe96}. We start with a brief history 
and elaborate on the underlying ideas. Subsequently we turn to 
biophysics, where generalized ensembles were introduced in 
Ref.~\cite{HaOk93,HaSc94}. Finally a scheme for biasing the 
Metropolis updating proposals \cite{Be03} is considered, which 
can be combined with generalized ensembles. 

As our emphasize is on explaining MCMC methods, we restrict ourselves 
to a simply models, which are well suited for illustrating the 
essence of a method. Our article is organized as follows. The next 
section introduces the MCMC method. Section~\ref{sec_AutoStat} 
discusses the statistical analysis of autocorrelated MCMC data.
Section~\ref{sec_GES} deals with generalized ensembles and 
section~\ref{sec_BMC} with biased MCMC updating. A short outlook
and conclusions are given in the final section~\ref{sec_final}.
It should be noted that these lecture notes are not supposed to 
be an unbiased review, but are based on work in which the author 
has been involved our is particularly interested.

\section{Markov Chain Monte Carlo}\label{sec_MCMC}

\subsection{Statistical preliminaries}\label{sec_Preliminaries}

Let $f(x)$ be a \index{probability density}probability density and 
$x^r$ its associated random variable. The 
\index{distribution function}{\bf (cumulative) 
distribution function}\index{cumulative distribution function} 
of the random variable $x^r$ is defined as 
\begin{equation} \label{df}
  F(x) = P(x^r\le x) = \int_{-\infty}^x f(x)\,dx
\end{equation}
where $P(x^r\le x)$ is the probability for $x^r\le x$. A particularly 
simple and important case is the \index{uniform distribution}{\bf 
uniform probability distribution} for random numbers between $[0,1)$, 
\begin{equation} \label{upd}
  u(x) = \cases{ 1~~{\rm for}~~0\le x < 1; \cr
                 0~~{\rm elsewhere}.  }
\end{equation}
Remarkably, the uniform distribution allows for the construction of 
general probability distributions.  Let
$$ y = F(x) = \int_{-\infty}^x f(x')\,dx' $$
and assume that the inverse $x=F^{-1}(y)$ exists. For $y^r$ being 
a uniformly distributed random variable in the range $[0,1)$ it
follows that
\begin{equation} \label{gd}
  x^r = F^{-1} (y^r)
\end{equation}
is distributed according to the probability density $f(x)$. To generate 
the uniform distribution on a computer, one relies on pseudo random 
number generators\index{random number generator}. 
Desirable properties are randomness according to 
statistical tests, a long period, computational efficiency, 
repeatability, portability, and homogeneity (all subsets of bits 
are random). Our purposes are served well by the generator of 
Marsaglia and collaborators \cite{MaZa90}, which comes as part
of the code of \cite{BBook}.

\subsection{Partition function and Potts models}
\label{sec_partition_function}

MC simulations of systems described by the
\index{Gibbs ensemble}Gibbs\index{canonical ensemble} canonical 
ensemble aim at calculating estimators of physical observables 
at a temperature $T$. In the following we choose units so that 
the \index{Boltzmann constant}Boltzmann 
constant becomes one, i.e. $\beta = 1/T$. Let us consider 
the calculation of the {\bf expectation value} of an {\bf observable} 
$\mathcal{O}$. Mathematically all systems on a computer are discrete, 
because a finite word length has to be used. Hence, the expectation 
value is given by the sum
\begin{eqnarray} \label{O}
 \widehat{\mathcal{O}} = \widehat{\mathcal{O}} (\beta) = 
 \langle \mathcal{O} \rangle &=& Z^{-1} \sum_{k=1}^K
 \mathcal{O}^{(k)}\,e^{-\beta\,E^{(k)} }\\
{\rm where}~~~
 Z\ =\ Z(\beta) &=& \sum_{k=1}^K e^{-\beta\,E^{(k)} }
\end{eqnarray}
is the \index{partition function}{\bf partition function}. The index 
$k=1,\dots , K$ labels the \index{configuration}{\bf configurations} 
of the system, and $E^{(k)}$ is the (internal) energy of configuration 
$k$. The configurations are also called \index{microstate}{\bf 
microstates}. To distinguish the configuration index from other 
indices, it is put in parenthesis.

We introduce generalized \index{Potts models}Potts models on 
$d$-dimensional hypercubic lattices with periodic boundary conditions 
(i.e., the models are defined on a torus in $d$ dimensions). 
Without being overly complicated, these models are general
enough to illustrate the essential features of MCMC simulations.
Various subcases are by themselves of physical interest. We define 
the energy function of the system by
\begin{equation} \label{E}
  E^{(k)} = - 2\,\sum_{\langle ij\rangle} \delta(q_i^{(k)},q_j^{(k)}) 
           + {2\,d\,N\over q} ~~{\rm where}~~ \delta (q_i,q_j)
 = \left\{ \begin{array}{c} 1\ {\rm for}\ q_i=q_j \\
         0\ {\rm for}\ q_i\ne q_j\,. \end{array}  \right.
\end{equation}
The sum $\langle ij\rangle$ is over the nearest neighbor lattice sites 
and $q_i^{(k)}$ is called the {\bf Potts spin} or {\bf Potts state} of 
configuration $k$ at site $i$. For the $q$-state Potts model, $q^{(k)}_i$ 
takes on the values $1,\dots ,q$. The case $q=2$ becomes equivalent to 
the \index{Ising model}Ising \index{ferromagnet}ferromagnet. See F.Y. 
Wu \cite{Wu82} for a review of Potts models. In $d=2$ dimensions the 
phase transition is second order for $q\le 4$ and first order for 
$q\ge 5$. The exact infinite volume latent heats 
\index{latent heat}$\triangle e_{s}$ and entropy discontinuities
\index{entropy discontinuities}$\triangle s$ were calculated by Baxter 
\cite{Ba73}, while the \index{interface tensions}interface tensions 
$f_s$ were derived later, see \cite{BoJa92} and references therein.

\subsection{Sampling, reweighting, and important configurations}

\begin{figure} \centering
\includegraphics[height=6cm]{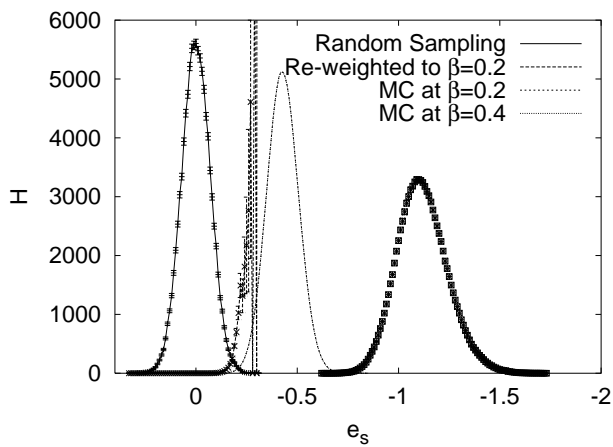}
\caption{Energy histograms of 100$\,$000 entries each for the Ising 
model on a $20\times 20$ lattice: Random Sampling gives statistically 
independent configurations at $\beta=0$. Histograms at $\beta=0.2$ 
and $\beta=0.4$ are generated with Markov chain MC. Reweighting of 
the $\beta=0$ random configurations to $\beta=0.2$ is shown to fail.
These are assignments {\tt a0301\_02} and {\tt a0303\_02} of 
\cite{BBook}.} \label{fig_2dI_h_es}
\end{figure}

For the Ising model (2-state Potts) it is straightforward to {\bf 
sample statistically independent configurations}. We simply have 
to generate $N$ spins, each either 0 or 1 with 50\% likelihood. 
This is called {\bf random sampling}. In Fig.~\ref{fig_2dI_h_es} 
a thus obtained histogram for the $2d$ Ising model {\bf energy 
per spin} is depicted. 

Note that it is important to distinguish the energy measurements 
on single configurations from the expectation value. The expectation 
value $\widehat{e}_{s}$ is a single number, while $e_{s}$ fluctuates.
From the measurement of many $e_{s}$ values one finds an estimator
of the mean, $\overline{e}_{s}$, which fluctuates with a reduced
variance.

The histogram entries at $\beta=0$ can be \index{reweighting}{\bf 
reweighted} so that they correspond to other $\beta$ values. We 
simply have to multiply the entry corresponding to energy $E$ by 
$\exp (-\beta E)$. Similarly histograms corresponding to the Gibbs 
ensemble at some value $\beta_0$ can be reweighted to other $\beta$ 
values. Care has to be taken to ensure that the involved arguments 
of the exponential function do not become too large~\cite{BBook}. 
Reweighting has a long history, which we discuss in 
section~\ref{sec_Rweight}.
 
In Fig.~\ref{fig_2dI_h_es} reweighting is done from $\beta_0=0$ to
$\beta=0.2$. But, by comparison to the histogram from a Metropolis 
MC calculation at $\beta=0.2$, the result is seen to be disastrous. 
The reason is easily identified: In the range where the $\beta=0.2$ 
histogram takes on its maximum, the $\beta=0$ histogram  has not a 
single entry. Our random sampling procedure misses the important 
configurations at $\beta =0.2$. Reweighting to new $\beta$ values 
works only in a range $\beta_0\pm \triangle \beta$, where 
$\triangle\beta\to 0$ in the infinite volume limit. Details
are given in section~\ref{sec_Rweight}.

Let us determine the important contributions to the partition function. 
The partition function can be re-written as a sum over energies
\begin{equation} \label{Z_sum_E}
 Z = Z(\beta) = \sum_E n(E)\, e^{-\beta\,E}
\end{equation}
where the unnormalized \index{spectral density}spectral density $n(E)$ 
is defined as the number of microstates $k$ with energy $E$. For a 
fixed value of $\beta$ the energy probability density
\begin{equation} \label{P_E}
 P_{\beta} (E) = c_{\beta}\, n(E)\, e^{-\beta E} 
\end{equation}
is peaked around the average value $\widehat{E}(\beta)$, where 
$c_{\beta}$ is a normalization constant determined by
$\sum_E P_{\beta}(E)=1$.  The important 
configurations at temperature $T=1/\beta$ are at the energy values for
which the probability density $P_{\beta}(E)$ is large. To sample them 
efficiently, one needs a procedure which generates the configurations 
with their \index{Boltzmann weights}Boltzmann \index{weights}weights
\begin{equation} \label{w_B}
 w_B^{(k)} = e^{-\beta E^{(k)}}\ . 
\end{equation}
The number of configurations $n(E)$ and the weights combine then 
so that the probability to generate a configuration at energy $E$
becomes $P_{\beta}(E)$ as given by equation~(\ref{P_E}).

\subsection{Importance sampling and Markov chain Monte Carlo}

For the canonical ensemble \index{importance sampling}{\bf importance 
sampling} generates configurations $k$ with probability
\begin{equation} \label{P_B}
 P_B^{(k)} = c_B\, w^{(k)}_B = c_B\, e^{-\beta E^{(k)}}
\end{equation}
where the constant $c_B$ is determined by the normalization condition
$\sum_k P_B^{(k)} = 1$. The vector $(P_B^{(k)})$ is called 
\index{Boltzmann state}{\bf Boltzmann state}. When configurations 
are stochastically generated with probability $P_B^{(k)}$, the 
expectation value becomes the {\bf arithmetic average:}
\begin{equation} \label{O_B}
  {\widehat {\cal O}} = {\widehat {\cal O}} (\beta) =
  \langle {\cal O} \rangle  = \lim_{N_K\to \infty} 
  {1\over N_K} \sum_{n=1}^{N_K} {\cal O}^{(k_n)}\ .
\end{equation}
Truncating the sum at some finite value of $N_K$, we obtain an 
\index{estimator}\index{expectation value}{\bf estimator of the 
expectation value}
\begin{equation} \label{estimate_O_B}
  {\overline {\cal O}} =
  {1\over N_K} \sum_{n=1}^{N_K} {\cal O}^{(k_n)}\ .
\end{equation}
Normally, we cannot generate configurations $k$ directly with the
probability~(\ref{P_B}), but they may be found as members
of the equilibrium distribution of a dynamic process. A 
\index{Markov process}{\bf Markov process} is a particularly
simple dynamic process, which generates configuration $k_{n+1}$
stochastically from configuration $k_n$, so that no information
about previous configurations $k_{n-1}, k_{n-2}, \dots$ is needed.
The elements of the Markov process
\index{time series}{\bf time series} are the configurations.
Assume that the configuration $k$ is given. Let the 
\index{transition probability}{transition probability}
to create the configuration $l$ in one step from $k$ be 
given by $W^{(l)(k)} = W[k\to l]$. The 
\index{transition matrix}{\bf transition matrix}
\begin{equation} \label{W}
 W = \left( W^{(l)(k)}\right)
\end{equation}
defines the Markov process. Note, that this matrix is a very big (never 
stored in the computer), because its labels are the configurations. To 
generate configurations with the desired probabilities, the matrix $W$ 
needs to satisfy the following properties:

\begin{description}

\item{(i)} \index{ergodicity}{\bf Ergodicity}:
\begin{equation} \label{ergodicity}
  e^{-\beta E^{(k)}} > 0 ~~{\rm and}~~ e^{-\beta E^{(l)}} > 0
  ~~~{\rm imply:}
\end{equation}
an integer number $n>0$ exists so that $(W^n)^{(l)(k)}>0$ holds.

\item{(ii)} {\bf Normalization}:
\begin{equation} \label{normalization}
 \sum_l W^{(l)(k)} = 1\ .
\end{equation}

\item{(iii)} \index{balance}{\bf Balance}: 
\begin{equation} \label{balance}
 \sum_k W^{(l)(k)}\, e^{-\beta E^{(k)}}\ =\ e^{-\beta E^{(l)}}\ .
\end{equation}
The Boltzmann state~(\ref{P_B}) is an eigenvector with eigenvalue $1$ 
of the matrix $W = (W^{(l)(k)})$.

\end{description}

An \index{ensemble}{\bf ensemble} is a collection of configurations for 
which to each configuration $k$ a probability $P^{(k)}$ is assigned, 
$\sum_k P^{(k)} = 1$. The \index{Gibbs ensemble}{\bf Gibbs or Boltzmann 
ensemble}\index{Boltzmann ensemble} $E_B$ is defined to be the ensemble 
with the probability distribution~(\ref{P_B}). 

An \index{equilibrium ensemble}{\bf equilibrium ensemble} $E_{eq}$ of 
the Markov process is defined by its probability distribution $P_{eq}$ 
satisfying
\begin{equation} \label{E_eq}
 W\, P_{eq} = P_{eq}\, , ~~{\rm in\ components}~~
 P_{eq}^{(l)}=\sum_k W^{(l)(k)} P_{eq}^{(k)}\ .
\end{equation}

{\bf Statement:} Under the conditions (i), (ii) and (iii) the Boltzmann 
ensemble is the {\bf only} equilibrium ensemble of the Markov process 
and an \index{fixed point}{\bf attractive fixed point}. Applying the 
transition matrix $n$ times give rise to an ensemble $E^n$. For 
$n\to\infty$ the distance between $E^n$ and the Boltzmann ensemble 
decreases asymptotically like\index{convergence of MCMC}
\begin{equation} \label{MCMC_convergence}
  ||E^n-E_B|| \le \exp (-\lambda\,n)\,||E^0-E_B||
\end{equation}
where $E^0$ is the initial ensemble and $\lambda > 0$ a constant.

For a proof the readers are referred to~\cite{BBook}. There are 
many ways to construct a Markov process satisfying (i), (ii)
and (iii).  A stronger condition than balance~(\ref{balance}) is

\begin{description}

\item{(iii')} \index{detailed balance}{\bf Detailed balance}:
\begin{equation} \label{detailed_balance}
 W^{(l)(k)}\, e^{-\beta E^{(k)}}\ =\ W^{(k)(l)} e^{-\beta E^{(l)}}\ .
\end{equation}
Using the normalization $\sum_k W^{(k)(l)} = 1$ detailed
balance implies balance.

\end{description}

At this point we have succeeded to replace the canonical ensemble
average by a time average over an artificial dynamics. Calculating
averages over large times, like one does in real experiments,
is equivalent to calculating averages of the ensemble. 

\subsection{Metropolis and heatbath algorithm for Potts models}

The \index{Metropolis algorithm|textbf}{\bf Metropolis algorithm} 
can be used whenever one knows how to calculate the energy of a 
configuration. Given a configuration $k$, the Metropolis algorithm 
proposes a configuration $l$ with probability
\begin{equation} \label{Metropolis_f}
 f(l,k) ~~{\rm normalized\ to}~~~ \sum_l f(l,k)=1\ .
\end{equation}
The new configuration $l$ is accepted with probability
\begin{equation} \label{Metropolis}
 w^{(l)(k)} = \min \left[ 1,\, {P_B^{(l)}\over P_B^{(k)}} \right] 
  = \cases{ 1~~{\rm for}~~E^{(l)}<E^{(k)} \cr
    \exp[-\beta (E^{(l)}-E^{(k)})]~~{\rm for}~~E^{(l)}>E^{(k)}.}
\end{equation}
If the new configuration is rejected, the old configuration has to 
be counted again. The \index{acceptance rate}{\bf acceptance rate} 
is defined as the ratio of accepted changes 
over proposed moves. With this convention we do not count a move as 
accepted when it proposes the at hand configuration.

The Metropolis procedure gives rise to the transition probabilities 
\begin{eqnarray} \label{Metropolis_W_ne}
 W^{(l)(k)} &=& f(l,k)\, w^{(l)(k)}\ \ {\rm for}\ \ l \ne k\\
{\rm and}~~
 W^{(k)(k)}& =& f(k,k) + \sum_{l\ne k} f(l,k)\, (1-w^{(l)(k)})\ .
\end{eqnarray}
Therefore, the ratio $\left( W^{(l)(k)}/W^{(k)(l)} \right)$ satisfies 
detailed balance~(\ref{detailed_balance}) if
\begin{equation} \label{symmetry}
 f(l,k)\ =\ f(k,l) ~~{\rm holds}\,.
\end{equation}
Otherwise the probability density $f(l,k)$ is unconstrained. So there 
is an amazing flexibility in the choice of the transition probabilities 
$W^{(l)(k)}$.  Also, the algorithm generalizes 
immediately to arbitrary weights.

If sites are chosen with the uniform probability distribution $1/N$ 
per site, where $N$ is the total number of spins, it is obvious that 
the algorithm fulfills detailed balance. It is noteworthy that the 
procedure remains valid when the spins are chosen in the systematic 
order $1,\dots ,N$.  Balance~(\ref{balance}) still holds, whereas 
detailed balance~(\ref{detailed_balance}) is violated (an exercise 
of Ref.~\cite{BBook}). 

If one performs multiple hits with the Metropolis algorithm at the
same spin (\index{multi-hit Metropolis}{\bf multi-hit Metropolis
algorithm}), the local Boltzmann distribution defined by its nearest 
neighbors is approached for an increasing number of hits.
The \index{heatbath algorithm}{\bf heatbath algorithm (HBA)} 
chooses a state $q_i$ directly with the local Boltzmann distribution 
defined by its nearest neighbors. The state $q_i$ can take on one of 
the values $1,\dots,q$ and, with all 
other states set, determines a value of the energy function~(\ref{E}). 
We denote this energy by $E(q_i)$ and the Boltzmann probabilities are
\begin{equation} \label{Potts_B_weights}
 P_B(q_i)\ =\ {\rm const}\ e^{-\beta\, E(q_i)}
\end{equation}
where the constant is determined by the normalization condition
\begin{equation} \label{Potts_B_normalization}
 \sum_{q_i=1}^q P_B(q_i)\ =\ 1\ .
\end{equation}        
In equation~(\ref{Potts_B_weights}) we can define $E(q_i)$ to be just 
the contribution of the interaction of $q_i$ with its nearest neighbors 
to the total energy and absorb the other contributions into the overall 
constant. Here we give a generic HBA which works for arbitrary values 
of $q$ and $d$ (other solutions can be more efficient). We calculate 
the cumulative distribution function of the 
heatbath probabilities
\begin{equation} \label{prob_hb}
  P_{HB}(q_i) = \sum_{q'_i=1}^{\tt q_i} P_B(q'_i)\ .
\end{equation}
The normalization condition~(\ref{Potts_B_normalization}) implies
$P_{HB}(q)=1$. Comparison of these cumulative probabilities
with a uniform random number $x^r$ yields the heatbath update
$q_i\to q'_i$. Note that in the heatbath procedure the original value
$q_i^{\rm in}$ does not influence the selection of $q_i^{\rm new}$. 

\subsection{Heatbath algorithm for a continuous system}

We give the $O(3)\ \sigma$ model as an example of a model with a 
continuous energy function.  Expectation values are calculated with 
respect to the partition function 
\begin{equation} \label{O3_Z}
  Z\ =\ \int \prod_i ds_i\ e^{-\beta E(\lbrace s_i \rbrace )}~~~
  {\rm of\ spins}~~\vec{s}_i = \left( \matrix{
                    s_{i,1}\cr s_{i,2}\cr s_{i,3}} \right)
\end{equation}
which are normalized to lie on the unit sphere, $(\vec{s}_i)^2=1$.
The measure $ds_i$ is defined by
\begin{equation} \label{O3_measure}
 \int ds_i\ =\ {1\over 4\pi} \int_{-1}^{+1} d \cos (\theta_i) 
                              \int_0^{2\pi}  d \phi_i\ , 
\end{equation}
where the polar ($\theta_i$) and azimuth ($\phi_i$) angles define the 
spin $s_i$ on the unit sphere. The energy is
\begin{equation} \label{O3_energy}
 E\ =\ - \sum_{\langle ij\rangle} \vec{s}_i \vec{s}_j\ ,  
\end{equation}
where the sum goes over the nearest neighbor sites of the lattice and
$\vec{s}_i \vec{s}_j$ is the dot product of the vectors. The $2d$ 
version of the model is of interest to field theorists because of its 
analogies with the four-dimensional Yang-Mills theory. In statistical 
physics the $d$-dimensional model is known as 
the\index{O(3) $\sigma$-model}\index{sigma model}
\index{Heisenberg ferromagnet}{\bf Heisenberg ferromagnet}
(references can be found in \cite{BBook}). 

We would like to update a single spin $\vec{s}$. The sum of its $2d$ 
neighbors is
$$\vec{S} = \vec{s}_1+\vec{s}_2+\dots +\vec{s}_{2d-1}+\vec{s}_{2d}\ .$$
Hence, the contribution of spin $\vec{s}$ to the energy is
$2d-\vec{s} \vec{S}$. We may propose a new spin $\vec{s}^{\,'}$ with 
the measure~(\ref{O3_measure}) by drawing two uniformly distributed 
random numbers
\begin{eqnarray} \nonumber
  \phi^r &\in & [0,2\pi) ~~~{\rm for\ the\ azimuth\ angle\ and} \\
  \cos (\theta^r) &=& x^r \in [-1,+1) ~~~{\rm for\ the\ cosine\ of\
  the\ polar\ angle}\,. \nonumber
\end{eqnarray}
This defines the probability function $f(\vec{s}^{\,'},\vec{s})$ of 
the Metropolis process, which accepts the proposed spin
$\vec{s}^{\,'}$ with probability 
$$ w(\vec{s}\to \vec{s}^{\,'})\ =\ \cases{
          1~~{\rm for}~~\vec{S}\vec{s}^{\,'}>\vec{S} \vec{s}\,,\cr
          \exp[-\beta(\vec{S}\vec{s}-\vec{S}\vec{s}^{\,'})]~~ 
          {\rm for}~~\vec{S}\vec{s}^{\,'}<\vec{S}\vec{s}\,.} $$

One would prefer to choose $\vec{s}^{\,'}$ directly with the
probability
$$ W(\vec{s}\to \vec{s}^{\,'})\ =\ P(\vec{s}^{\,'};\vec{S})\ =\
  {\rm const}\, e^{\beta\, \vec{s}^{\,'} \vec{S}}\ . $$
The \index{heatbath algorithm}HBA creates this distribution.  
Implementation of it becomes feasible when the energy function 
allows for an explicit calculation of the probability 
$P(\vec{s}^{\,'};\vec{S})$. This is an easy task for the 
$O(3)$ $\sigma$-model. Let
$$ \alpha = {\rm angle} (\vec{s}^{\,'},\vec{S}),\ \
x=\cos (\alpha )\ \ {\rm and}\ \  S=\beta |\vec{S}|\ . $$
For $S=0$ a new spin $\vec{s}^{\,'}$ is simply obtained by random
sampling. We assume in the following $S>0$. 
The Boltzmann weight becomes $\exp (x S)$ and the normalization
constant follows from
$$ \int_{-1}^{+1} dx\, e^{x S}\ =\ {2\over S}\, \sinh (S)\ .$$
Therefore, the desired probability is
$$ P(\vec{s}^{\,'};\vec{S})\ =\  
  {S \over 2\sinh (S)}\, e^{x S}\ =: f(x) $$
and Eq.~(\ref{gd}) can be used to generate events with the probability 
density $f(x)$. A uniformly distributed random number $y^r\in [0,1)$ 
translates into
\begin{equation} \label{O3_heat_bath-xr}
  x^r = \cos{\alpha^r} = {1\over S} \ln\, [\, \exp (+S) -
  y^r\, \exp (+S) + y^r\, \exp (-S) ]\ . 
\end{equation}
To give $\vec{s}^{\,'}$ a direction in the plane orthogonal to 
$\vec{S}$, one chooses a uniformly distributed angle $\beta^r$
in the range $0\le\beta^r < 2\pi$. Then, $x^r=\cos{\alpha^r}$ and 
$\beta^r$ completely determine $\vec{s}^{\,'}$ with respect to 
$\vec{S}$. Before storing $\vec{s}^{\,'}$ in the computer memory, 
we have to calculate coordinates of $\vec{s}^{\,'}$ with respect to 
a Cartesian coordinate system, which is globally used for all spins 
of the lattice. This amounts to a linear transformation. 
 
\section{Statistical Errors of MCMC Data\label{sec_AutoStat}} 

\index{statistical errors}In large scale MC simulation it may take 
months, possibly years, to collect the necessary statistics. For such 
data a thorough error analysis is a must. A typical MC simulation 
falls into two parts: 

\begin{enumerate}

{\bf \item Equilibration:} Initial 
sweeps are performed to reach the equilibrium distribution. During 
these sweeps measurements are either not taken at all or they have 
to be discarded when calculating equilibrium expectation values.

{\bf \item Data Production:}
\index{data production}Sweeps with measurements are performed.
Equilibrium expectation values are calculated from this statistics.

\end{enumerate}

A rule of thumb is: {\bf Do not spend more than 50\% of your CPU time 
on measurements.} \index{error analysis}The reason for this rule is 
that one cannot be off by a factor worse than two ($\sqrt{2}$ in the 
statistical error).

How many sweeps should be discarded for reaching equilibrium? In 
some situations this question can be rigorously answered with the 
\index{coupling from the Past}{\it Coupling from the Past} method 
\cite{PrWi98} (for a review see~\cite{Ke05}). The next best thing 
to do is to measure the integrated autocorrelation time and to discard, 
after reaching a visually satisfactory situation, a number of sweeps 
which is larger than the integrated autocorrelation time. In practice 
even this can often not be achieved.

Therefore, it is re-assuring that it is sufficient to pick the number 
of discarded sweeps approximately right. With increasing statistics 
the contribution of the non-equilibrium data dies out like $1/N$, 
where $N$ is the number of measurements. This is eventually swallowed 
by the statistical error, which declines only like $1/\sqrt{N}$. The 
point of discarding the equilibrium configurations is that the factor 
in front of $1/N$ can be large.

There can be far more involved situations, like that the Markov 
chain ends up in a metastable configuration, which may even 
stay unnoticed (this tends to happen in complex systems like spin 
glasses or proteins).
 
\subsection{Autocorrelations} 

We like to estimate the expectation value $\widehat{f}$ of some physical
observable. We assume that the system has reached
equilibrium.  How many MC sweeps are needed to estimate
$\widehat{f}$ with some desired accuracy? To answer this question, one
has to understand the autocorrelations within the Markov chain.

Given is a {\bf time series} of $N$ measurements from a Markov process
\begin{equation} \label{time_series}
 f_i = f(x_i), ~~ i=1,\dots,N\ , 
\end{equation}
where $x_i$ are the configurations generated.
The label $i=1,\dots,N$ runs in the temporal 
order of the Markov chain and the elapsed time (measured in updates 
or sweeps) between subsequent measurements $f_i$, $f_{i+1}$ is always 
the same.  The estimator of the 
\index{expectation value}{expectation value} $\widehat{f}$ is
\begin{equation} \label{fhat_estimator}
 \overline{f} = {1\over N} \sum f_i\ .
\end{equation}
With the notation $$t=|i-j|$$ the definition of the 
\index{autocorrelations}{\bf autocorrelation function} 
of the observable $\widehat{f}$ is 
\begin{equation} \label{autocorrelation}
 \widehat{C}(t) = \widehat{C}_{ij} = \langle\, (f_i-\langle f_i\rangle)\, 
 (f_j - \langle f_j\rangle)\, \rangle =
 \langle f_i f_j\rangle - \langle f_i \rangle\, \langle f_j\rangle = 
 \langle f_0 f_t\rangle - \widehat{f}^{\,2} \end{equation}
where we used that translation invariance in time holds for the 
equilibrium ensemble. The asymptotic behavior for large $t$ is
\begin{equation} \label{tau_exp}
 \widehat{C}(t) \sim \exp \left( - {t\over \tau_{\rm exp}} \right)
 ~~{\rm for}~~ t \to \infty ,
\end{equation}
where $\tau_{\rm exp}$ is called\index{exponential autocorrelation time} 
{\bf exponential autocorrelation time} and is related to the second 
largest eigenvalue $\lambda_1$ of the transition matrix by 
$\tau_{\rm exp}=-\ln\lambda_1$
under the assumption that $f$ has a non-zero projection on the
corresponding eigenstate. Superselection rules are possible so that 
different autocorrelation times reign for different operators.

The variance of $f$ is a special case of the 
autocorrelations~(\ref{autocorrelation})
\begin{equation} \label{C0}
 \widehat{C} (0) = \sigma^2 (f)\ .
\end{equation}
Some algebra \cite{BBook} shows that the variance of the estimator 
$\overline{f}$~(\ref{fhat_estimator}) for the mean and the 
autocorrelation function~(\ref{autocorrelation}) are related by
\begin{equation} \label{variance_correlated}
 \sigma^2(\overline{f})\ =\ {\sigma^2(f)\over N} \left[ 1 +
 2 \sum_{t=1}^{N-1} \left( 1 -{t\over N} \right)\, \widehat{c}(t) \right]
 ~~{\rm with}~~ \widehat{c}(t) = {\widehat{C}(t)\over \widehat{C}(0)}\ .
\end{equation}
This equation ought to be compared with the corresponding equation 
for uncorrelated random variables:
$ \sigma^2(\overline{f}) = {\sigma^2(f)/ N}$. The difference is the 
factor in the bracket of~(\ref{variance_correlated}), which defines 
the\index{integrated autocorrelation time}
\index{autocorrelations}{\bf integrated autocorrelation time}
\begin{equation} \label{ia_time_N}
 \tau_{\rm int}\ =\ \left[ 1 + 2 \sum_{t=1}^{N-1} 
 \left(1-{t\over N}\right)\, \widehat{c}(t) \right]\ .
\end{equation}
For correlated data the variance of the mean is by the factor 
$\tau_{\rm int}$ larger than the corresponding \index{variance}variance 
for uncorrelated data. In most simulations one is interested in the 
limit $N\to\infty$ and equation~(\ref{ia_time_N}) becomes
\begin{equation} \label{ia_time}
\tau_{\rm int}\ =\ 1 + 2 \sum_{t=1}^{\infty} \widehat{c}(t)\ . 
\end{equation}
The numerical estimation of the integrated autocorrelation time 
faces difficulties. The variance of the estimator for (\ref{ia_time}) 
diverges, because for large $t$ each $\overline{c}(t)$ adds a constant 
amount of noise, whereas the signal dies out like 
$\exp(-t/\tau_{\rm exp})$. To obtain an estimate one 
considers the $t$-dependent estimator
\begin{equation} \label{iat_estimate}
\overline{\tau}_{\rm int}(t)\ =\ 1 + 2 \sum_{t'=1}^t
\overline{c}(t') 
\end{equation}
and looks out for a {\bf window} in $t$ for which 
$\overline{\tau}_{\rm int}(t)$ is flat. 

\subsection{Integrated autocorrelation time and binning}

Using \index{binning}binning (also called \index{blocking}blocking) 
the integrated autocorrelation time can also be estimated 
via the \index{variance ratio}variance ratio. We bin the time 
series~(\ref{time_series}) into $N_{bs}\le N$ bins of
\begin{equation} \label{Nb}
 N_b = {\tt NBIN} = \left[ N\over N_{bs} \right] =
 \left[ {\tt NDAT} \over {\tt NBINS} \right]
\end{equation}
data each. Here $[.]$ stands for Fortran integer division, {\it i.e.}, 
$N_b={\tt NBIN}$ is the largest integer $\le N/N_{bs}$, implying
$N_{bs}\cdot N_b\le N$. It is convenient to choose the values of $N$ 
and $N_{bs}$ so that $N$ is a multiple of $N_{bs}$. The binned data 
are the averages
\begin{equation} \label{binned_data}
  f^{N_b}_j = {1\over N_b} \sum_{i=1+(j-1)N_b}^{jN_b} f_i 
  ~~~~{\rm for}~~ j=1,\dots ,N_{bs}\ .
\end{equation}
For $N_b > \tau_{\rm exp}$ the autocorrelations are essentially reduced 
to those between nearest neighbor bins and even these approach zero 
under further increase of the binsize.

For a set of $N_{bs}$ binned data $f^{N_b}_j$, $(j=1,\dots ,N_{bs})$ we 
may calculate the mean with its naive error bar. Assuming for the 
moment an infinite time series, we find the integrated autocorrelation 
time~(\ref{ia_time_N}) from the following ratio of sample 
variances
\begin{equation} \label{iat_binning} 
  \tau_{\rm int}\ =\ \lim_{N_b\to\infty}\, \tau_{\rm int}^{N_b}\ 
  ~~{\rm with}~~\ \tau_{\rm int}^{N_b}\ =\ \left( 
  { s^2_{\overline{f}^{N_b}} \over s^2_{\overline{f}} }\right)\ .
\end{equation}
In practice the $N_b\to\infty$ limit will be reached for a sufficiently 
large, finite value of $N_b$. The statistical error of the 
$\tau_{\rm int}$ estimate~(\ref{iat_binning}) is, in the first 
approximation, determined by the errors of $s^2_{\overline{f}^{N_b}}$. 
The typical situation is then that, due to the central limit theorem, 
the binned data are approximately Gaussian, so that the {\bf error of 
$s^2_{\overline{f}^{N_b}}$ is analytically known} from the $\chi^2$ 
distribution. Finally, the fluctuations of 
$s^2_{\overline{f}}$ of the denominator give rise to a small
correction which can be worked out~\cite{BBook}. 

Numerically most accurate estimates of $\tau_{\rm int}$ are obtained 
for the finite binsize $N_b$ which is just large enough that the binned 
data~(\ref{binned_data}) are practically uncorrelated. 
While the Student distribution shows that the confidence intervals 
of the error bars from 16 uncorrelated normal data are reasonable 
approximations to those of the Gaussian standard deviation,
about 1000 independent data are needed to provide a decent 
estimate of the corresponding variance (at the 95\% confidence 
level with an accuracy of slightly better than 10\%). 
It makes sense to work with error bars from 16 binned data, but the 
error of the error bar, and hence a reliable estimate of 
$\tau_{\rm int}$, requires far more data.
 
\subsection{Comparison of MCMC algorithms}

\begin{figure} \centering
\includegraphics[height=6cm]{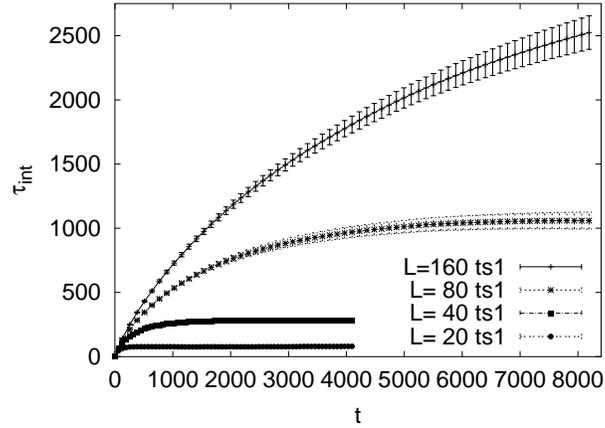}
\caption{One-hit Metropolis algorithm 
with sequential updating: Lattice size dependence of the integrated 
autocorrelation time for the $d=2$ Ising model at its critical 
temperature (assignment {\tt a0402\_02}~D of \cite{BBook}). The 
ordering of the curves is identical with the ordering of the labels 
in the figure.} \label{fig_iat2_large} 
\end{figure}

\begin{figure} \centering
\includegraphics[height=6cm]{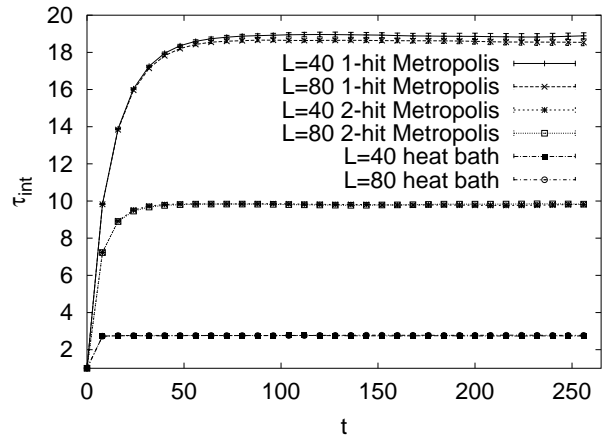}
\caption{Systematic updating: Comparison of the integrated 
autocorrelation times of the 1-hit and 2-hit Metropolis algorithms 
and the heat bath algorithm for the 10-state Potts model on $L\times L$ 
lattices at $\beta =0.62$ (assignment {\tt a0402\_06}). The $L=40$ 
and $L=80$ curves lie almost on top of one another.} \label{fig_iathb}
\end{figure}

Figure~\ref{fig_iat2_large} illustrates $2d$ Ising model simulations on 
the critical point of its second order phase transition, 
$\beta=\beta_c=\ln(1+\sqrt{2})/2$. 
\index{critical slowing down}{\bf Critical slowing down} is
observed: An increase $\tau_{\rm int}\sim L^z$ with lattice
size, where $z\approx 2.17$ is the 
\index{dynamical critical exponent}{\bf dynamical critical exponent}
of the $2d$ Ising model. Estimates of $z$ are compiled in~\cite{LaBiBook}.
Using another MC dynamics the critical slowing down can be overcome
by cluster updating \cite{SwWa87,Wo89}.

Figure~\ref{fig_iathb} exhibits the {\bf improvements of heat 
bath over Metropolis updating}\index{heatbath algorithm} for the 10-state 
$d=2$ Potts model at $\beta=0.62$. For this first order phase 
transition there is practically no lattice size dependence of the 
integrated autocorrelation time, once the lattices are large enough. 
We see that the 2-hist Metropolis updating reduces $\tau_{\rm int}$ 
by about a factor of two and the heatbath updating reduces it by
about a factor of five.

\subsection{Jackknife estimators} 

Often one wants to estimate a non-linear function of the mean
${\widehat x}$ of some statistical variables ${\widehat f} = 
f({\widehat x})$ where the estimator of $\widehat x$ and 
${\widehat f}$ are
\begin{equation} 
  {\overline x} = {1\over N} \sum_{i=1}^N x_i\,,\qquad 
  {\overline f} = f({\overline x})\ .
\end{equation}
Typically, the bias is of order $1/N$:
\begin{equation} \label{f_bias}
 {\rm bias}\ ({\overline f})\ =\ {\widehat f} -  
 \langle {\overline f}\rangle\ =\ {a_1 \over N} + {a_2 \over N^2} 
 + O({1\over N^3})
\end{equation}
where $a_1$ and $a_2$ are constants. But for the biased estimator we 
lost the ability to estimate the variance $\sigma^2 ({\overline f}) 
= \sigma^2 (f) / N$ via the standard equation
\begin{equation} \label{f_bad_variance}
 s^2 ({\overline f})\ =\ {1\over N} s^2 (f)\ =\
 {1\over N\, (N-1)} \sum_{i=1}^N (f_i - {\overline f})^2\ ,  
\end{equation}
because $f_i=f(x_i)$ is not a valid estimator of ${\widehat f}$. The 
error bar problem for the estimator $\overline{f}$ is conveniently 
overcome by using \index{jackknife estimators}{\bf jackknife 
estimators} ${\overline f^J}$, $f_i^J$, defined by
\begin{equation} \label{jackknife_estimators}
 {\overline f^J}\ =\ {1\over N} \sum_{i=1}^N f^J_i
 ~~{\rm with}~~ f_i^J\ =\ f(x_i^J) ~~{\rm and}~~ 
   x_i^J\ =\ {1\over N-1} \sum_{k\ne i} x_k\ . 
\end{equation}
The estimator for the variance $\sigma^2 ({\overline f^J})$ is
\begin{equation} \label{jackknife_variance}
 s^2_J ({\overline f^J})\ =\ {N-1 \over N} 
 \sum_{i=1}^N (f^J_i - {\overline f^J})^2\ . 
\end{equation}
Straightforward algebra shows that in the unbiased case the estimator 
of the jackknife variance~(\ref{jackknife_variance}) reduces to the 
normal variance~(\ref{f_bad_variance}). Notably, only order $N$ (not 
$N^2$) operations are needed to construct the jackknife averages 
$x_i^J,\ i=1,\dots ,N$ from the orginal data.

The jackknife method was introduced in the 1950s \cite{Qu56,Tu58}. For 
a review see~\cite{BBook}). It is recommended as the standard for error 
bar calculations of biased estimators.
 
\subsection{Self-consistent versus reasonable error analysis}

By visual inspection of the time series, one may get an impression about 
the length of the out-of-equilibrium part of the simulation. On top of 
this one should still choose
\begin{equation} \label{auto_equi}
{\tt nequi}\ \gg\ \tau_{\rm int}\ ,
\end{equation}
to allow the system to settle. That is a first reason, why it appears 
necessary to control the integrated autocorrelation time of a MC 
simulations. A second reason is that we have to control the error 
bars of the equilibrium part of our simulation. Ideally the error 
bars are calculated as
\begin{equation} \label{error_best}
\triangle \overline{f}\ =\ \sqrt{\sigma^2(\overline{f})}
~~~{\rm with}~~~
\sigma^2(\overline{f})\ =\ \tau_{\rm int}\,{\sigma^2(f)\over N}\ .
\end{equation}
This constitutes a \index{error analysis}{\bf self-consistent error 
analysis} of a MC simulation.

However, the calculation of the integrated autocorrelation time may
be out of reach.  Many more than the about twenty independent data are 
needed, which according to the Student distribution are sufficient to 
estimate mean values with reasonably reliable error bars.

In practice, one has to be content with what can be done. {\bf Often 
this means to rely on the binning method.} We simply calculate error 
bars of our ever increasing statistics with respect to a fixed number of 
\begin{equation} \label{NBINS_choice}
{\tt NBINS}\ \ge\ 16\ .
\end{equation}
In addition, we may put 10\% of the 
initially planned simulation time away for reaching equilibrium. 
{\it A-posteriori}, this can always be increased. Once the statistics 
is large enough, our small number of binned data become effectively 
independent and our error analysis is justified. 

How do we know that the statistics has become large enough? In practical 
applications there can be indirect arguments, like FSS estimates, which 
tell us that the integrated autocorrelation time is in fact (much) 
smaller than the achieved bin length. This is no longer self-consistent, 
as we perform no explicit measurement of $\tau_{\rm int}$, but it is a 
\index{error bar}{\bf reasonable error analysis}.

\section{Generalized Ensembles for MCMC Simulations}\label{sec_GES}

The MCMC schemes discussed so far simulate the Gibbs canonical 
ensemble. Mean values of physical observables at the temperature 
chosen are obtained as arithmetic averages of the measurements.  
However, in statistical physics one would like to control the 
partition function, which allows to calculate observables at all 
temperatures and for the the proper normalization of the entropy 
and free energy. Also configurations, which are rare in the canonical, 
but well represented in another ensemble can be of physical interest.
Finally the efficiency of the Markov process, i.e., the computer time
needed for the convergence of an estimator of a physical quantity
to a desired accuracy can depend greatly on the ensemble in which
the simulations are performed.

\subsection{Reweighting of the canonical ensemble}\label{sec_Rweight}

A first attempt to calculate the partition function by MCMC simulations 
dates back to a 1959 paper by Salsburg et al.~\cite{Sa59}. As was 
already noticed by the authors their method  is restricted to very 
small lattices. The reason is that their approach relies on what one 
calls in the modern language \index{reweighting}{\bf reweighting} of 
the Gibbs canonical ensemble. It extrapolates data from a canonical 
simulation at one temperature to the canonical ensemble at another 
temperature.  

The reweighting method has a long history. McDonald and Singer
\cite{McSi67} were the first to use the equations of \cite{Sa59} to 
evaluate physical quantities over a range of temperatures. But 
thereafter the method was essentially forgotten and a recovery in 
lattice gauge theory \cite{FaMa82,Ma84} focused on calculating complex 
zeros of the partition function. It remained to the paper by Ferrenberg 
and Swendsen \cite{FeSw88}, to formulate crystal-clear for what the 
method is particularly good, and for what not: It allows to focus 
on peaks of appropriate observables in the thermodynamic scaling limit, 
but it does not allow to cover a finite temperature range in the 
infinite volume limit. Off critical points, the reweighting range 
$\triangle\beta$ in $\beta=1/(kT)$ decreases like $\triangle\beta 
\sim 1/\sqrt{N}$, where $N$ is the number of degrees of freedom, which 
parametrizes the size of the system (e.g., the number of atoms). This 
follows from the fact that the energy is an extensive thermodynamic 
quantity, $E \sim N$ with fluctuations $\sim \sqrt{N}$. As $\beta$ 
multiplies the energy, the change stays within the fluctuations as 
long as $\triangle\beta\,N \sim \sqrt{N}$, so that $\triangle\beta 
\sim 1/\sqrt{N}$ follows. 

At second order phase 
transitions the reweighting range actually increases, because 
critical fluctuations are larger than non-critical fluctuations. 
One has then $\triangle E \sim N^x$ with $1/2<x<1$ and the requirement 
$\triangle\beta\,N \sim N^x$ yields $\triangle\beta \sim N^{x-1}$. 
For first order phase transitions one has a latent heat $\triangle E
\sim N$, but this does not mean that the reweighting range becomes of 
order one. In essence, the fluctuations collapse, because the two 
phases become separated by an interface. One is back to fluctuations 
within either of the two phases where $\triangle\beta\sim 1/\sqrt{N}$
holds.

To estimate the partition function over a finite range $\triangle e$ in 
the energy density $e=E/N$, i.e., $\triangle E \sim N$, one can patch
the histograms from canonical simulations at several temperatures.
Such \index{multi-histogram methods}{\bf multi-histogram methods} have 
also a long tradition too. In 1972 Valleau and Card \cite{VaCa72} 
proposed the use of overlapping bridging distributions and called 
their method \index{multistage sampling}``multistage sampling''. 
Free energy and entropy calculations become possible when one can 
link the temperature region of interest with a point in configuration 
space for which exact values of these quantities are known. 
Ref.~\cite{FeSw88} stimulated a renaissance of this approach. Modern 
work \cite{FeSw89,AlBe90,AlBe92} developed efficient techniques
to combine the overlapping distributions into one estimate of the 
spectral density and to control the statistical errors of the estimate. 
However, the patching of histograms from canonical simulations faces a 
number of limitations:

\begin{enumerate}

\item The number of canonical simulations diverges like $\sqrt{N}$
when one wants to cover a finite, non-critical range of the energy
density.

\item At first order phase transition point, the canonical probability
of configuration with an interface decreases $\sim \exp(-f_s\,A)$. Here 
$f_s$ is the interfacial surface tension and $A$ the minimal area of an 
interface, which divides the system into subsystems of distinct phases. 
For a system of volume $L^d$ the area $A$ diverges $\sim L^{d-1}$ in
the infinite volume limit $L\to\infty$.

\end{enumerate}

\subsection{Generalized ensembles}

One can cope with the difficulties of multi-histogram methods
by allowing arbitrary sampling
distributions instead of just the Gibbs-Boltzmann ensemble. This 
was first recognized by Torrie and Valleau~\cite{ToVa77} when they 
introduced \index{umbrella sampling}{\bf umbrella sampling}. However, 
for the next thirteen years application of this idea remained mainly 
confined to physical chemistry. That there is a potentially very broad 
range of applications for the basic idea remained unrecognized. A major 
barrier, which prevented researchers from trying such extensions, was 
certainly the apparent lack of direct and straightforward ways of 
determining suitable weighting functions for problems at hand. In the 
words of Li and Scheraga~\cite{LiSc88}: {\it The difficulty of finding 
such weighting factors has prevented wide applications of the umbrella 
sampling method to many physical systems.}

The turn-around came with the introduction of the {\bf multicanonical 
ensemble}\index{multicanonical ensemble} \cite{BeNe91,BeNe92,Be92}. 
These papers focused on one well-defined weight function, up to a 
normalization constant the inverse spectral density, 
\begin{equation} \label{w_muca}
  w_{\rm muca}(E)\sim 1/n(E)~~{\rm for}~~E_{\min}\le E\le E_{\max}\,,
\end{equation}
where $n(E)$ is the number of states, and offered a variety of methods 
to find a \index{working approximation}{\bf ``working approximation''}
of the weight function. Here a working approximation is defined as 
being accurate enough, so that the desired energy range will indeed 
be covered after the weight factors are fixed. A typical 
multicanonical simulation consists then of three parts:

\begin{enumerate}

\item Construct a working approximation of the weight function
      $w_{\rm muca}$.

\item Perform a conventional MCMC simulation with these weight
      factors.

\item Reweight the data to the Gibbs-Boltzmann ensemble at desired
      temperatures to obtain estimates of canonical expectation values 
      for observables of physical of interest.

\end{enumerate}

For instance, for the statistical physics system considered in
\cite{BeNe92}, the $2d$ 10-state Potts model, finite size scaling
consideration allow to construct the weight function on a larger 
lattice in one step from the information about the spectral density
on the already simulated smaller lattices. This is a solution to
step (1) in this particular case. The simulation and data analysis
(see \cite{BBook} for details) is then rather straightforward.
Still, such conceptual simplifications might have changed little on
the situation that non-canonical ensembles were rarely used, if 
there were not other favorable circumstance.

One was that the paper by \cite{BeNe92} estimated the interfacial 
tension of the $2d$ 10-state Potts model and produced a result, which 
was an order of magnitude smaller than previous numerical estimates 
by renown computational scientists. Normally this would have been 
followed by an extended debate of the advantages and disadvantages 
of the two competing messages. However, shortly after publication of 
the numerical estimates it turned out that the interfacial tension 
of the $2d$ 10-state Potts models can be calculated analytically 
\cite{BoJa92}, and the multicanonical estimate was within 3\% of 
the rigorous result. This attracted attention and gave a lot of 
researchers confidence in the method.

Another phenomenon was, that at $1991 \pm 5$ years a number of papers 
\cite{SwWa86,Ge91,MaPa92,LyMa92,HuNe96} emerged in the literature, 
which all aimed at improving MCMC calculations by extending the 
confines of the canonical ensemble. Most important has been the
\index{replica exchange method}{\bf replica exchange method}, which 
is also known under the names \index{parallel tempering}{\bf parallel 
tempering}\index{tempering} and \index{multiple Markov chains}{\bf 
multiple Markov chains}. In the context of spin glass simulations an 
exchange of partial lattice configurations at different temperatures
was proposed by Swendsen and Wang \cite{SwWa86}. But it was only in 
the later works \cite{Ge91,HuNe96}, essentially by rediscovery, 
recognized that the special case for which entire configurations 
at distinct temperatures are exchanged is of utmost importance. 

In the successful replica exchange method one performs $n$ canonical 
MC simulations at different $\beta$-values with Boltzmann weight factors
\begin{equation} \label{PT_weights}
  w_{B,i} (E^{(k)}) = e^{-\beta_i E^{(k)}} = e^{-H}\,,\
  i=0, \dots , n-1
\end{equation}
where $\beta_0 < \beta_1 < ... < \beta_{n-2} < \beta_{n-1}$, and
exchanges neighboring $\beta$-values
\begin{equation} \label{ex_beta_PT}
  \beta_{i-1} \longleftrightarrow \beta_i\ ~~{\rm for}~~
  i=1, \dots , n-1\ .
\end{equation}
Their joint Boltzmann factor is
\begin{equation} \label{Boltz_2}
  e^{-\beta_{i-1}E_{i-1}-\beta_iE_i} =  e^{-H}
\end{equation}
and the $\beta_{i-1}\leftrightarrow\beta_i$ exchange leads to 
\begin{eqnarray} \label{delH_PT}
  - \triangle H & = & \left( -\beta_{i-1} E^{(k)}_i -   
  \beta_i E^{(k')}_{i-1} \right) - \left( -\beta_i E^{(k)}_i
  -   \beta_{i-1} E^{(k')}_{i-1} \right) \\ \nonumber
  & = &\ \left( \beta_i-\beta_{i-1} \right)\,
      \left( E^{(k)}_i - E^{(k')}_{i-1} \right)
\end{eqnarray}
which is accepted or rejected according to the Metropolis algorithm,
{\it i.e.}, with probability one for $\triangle H\le 0$ and with 
probability $\exp (-\triangle H)$ for $\triangle H>0$.
The $\beta_i$-values should to be determined so that an appropriate 
large acceptance rate is obtained for {\bf each} $\beta$ exchange. 
This can be done by recursions~\cite{BBook}, which are straightforward 
modifications of a method introduced in Ref.~\cite{KeRe94}. 

Finally, and perhaps most important: From about 1992 on applications 
of generalized ensemble methods diversified tremendously. This is 
documented in a number of reviews \cite{Ja98,HaOk99,Be00,MiSu01}. In 
the next section we focus on the use of generalized ensembles in 
biophysics.

\subsection{Generalized ensembles and biophysics}\label{sec_GEBio}

In Ref.~\cite{BeCe92} the \index{multicanonical ensemble}multicanonical 
ensemble was first used for simulations of 
\index{complex systems}complex systems with frustrated 
interactions\index{frustrated interactions}, in that case 
the Edwards-Anderson Ising spin glass. Multicanonical 
simulations of biologically relevant molecules followed
\cite{HaOk93,HaSc94}, in Ref.~\cite{HaSc94} under the name
\index{entropic sampling}{``entropic sampling''}, but this is
just identical with multicanonical sampling~\cite{BeHa95}.

The interactions defined by an energy function are frustrated if one 
cannot simultaneously align all variables favorably with respect to 
their mutual interactions. So one gets ``frustrated'', a situation 
well known to political decision makers. In physical models 
frustrated interactions lead to a rugged free energy landscape.  
\index{rugged free energy landscape}Canonical MCMC simulations
tend to get stuck in the \index{free energy barriers}free energy 
barriers. In a nutshell, another ensemble may smoothen out such
barriers, jump them, or at least allow to escape from them, for 
instance through a disordered phase. This can accelerate the over-all 
convergence of the MCMC process considerably. If all parameters 
are chosen right, reweighting\index{reweighting} will finally allow 
to reconstruct canonical ensemble expectation values at desired 
temperatures.

The parallel tempering (replica exchange) method was introduced in 
Ref.~\cite{Ha97} to the simulation of biomolecules. In particular its 
extension to Molecular Dynamics \cite{SuOk99} has subsequently been 
tremendously successful. Nowadays folding of small proteins is 
achieved using PC clusters and it appears that all these simulations 
rely on some form of the replica exchange method.

\subsection{Overcoming free energy barriers}

\begin{figure} \centering
\includegraphics[height=6cm]{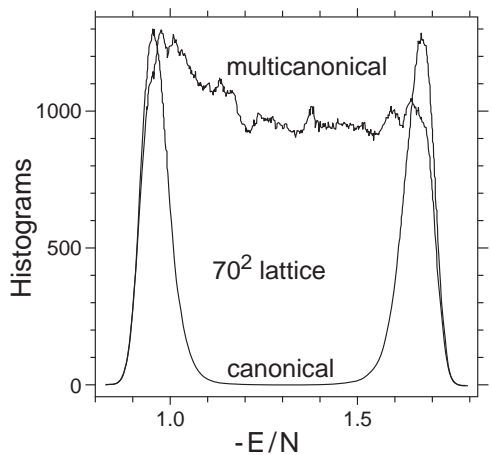}
\caption{Multicanonical $P_{mu}(E)$ together with canonical $P(E)$ 
energy distribution as obtained in Ref.\cite{BeNe92} for the $2d$ 
10-state Potts model on a $70\times 70$ lattice.} \label{fig_2d10q}
\end{figure}

Basic mechanisms for overcoming energy barriers are best illustrated 
for first-order phase transitions. There one deals with the simplified 
situation of a single barrier, which is easily visualized by plotting
histograms of an appropriate observable. To give a first example, 
Fig.~\ref{fig_2d10q} shows for the $2d$ 10-state Potts model the
canonical energy histogram at a pseudo-critical temperature versus 
the energy histogram of a multicanonical simulation on a $70\times 70$
lattice \cite{BeNe92}. The energy barrier is overcome by enhancing the 
probabilities for configurations, which are suppressed in the canonical 
ensemble due to an interfacial tension.

\begin{figure} \centering
\includegraphics[height=6cm]{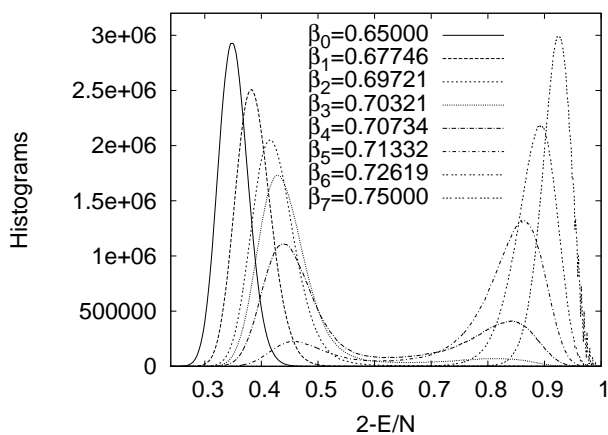}
\caption{Energy histograms from a parallel tempering simulation with 
eight processes for the $2d$ 10-state Potts model on $20\times 20$ 
lattices (assignment {\tt a0603\_04} in~\cite{BBook}). 
\label{fig_pt_hist} }
\end{figure}

The same barrier can be overcome by a\index{parallel tempering} 
parallel tempering simulation, but in a quite different way.
Fig.~\ref{fig_pt_hist} shows the histograms from a parallel
tempering simulation with eight processes on $20\times 20$
lattices. Each histogram corresponds to a fixed temperature, 
given by the $\beta$ values in the figure. The $\beta_4$ and 
$\beta_5$ values produce the clearest double peak histograms. 
For $\beta_4$ the higher peak is in the disordered region and for 
$\beta_5$ it is in the ordered region. So the barrier can be
``jumped'' when there are at least two temperatures in the ensemble, 
which are sufficiently close to the particular pseudo-critical 
temperature for which the two peaks of the histogram assume equal 
heights\index{equal heights/weights} (pseudocritical temperatures 
may also be defined by giving equal weights to the two parts of the 
histogram~\cite{BoKa92}). One of these two temperatures has to be 
in the ordered, the other in the disordered phase, and their start 
configurations have to be in the corresponding phases. The barrier 
can be jumped by an exchange of these two temperatures. If the barrier 
is high enough, so that a single canonical simulations does not cross
it during the simulation time, the jumping process determines the 
relative height of the two barriers. Adding more close-by temperatures 
to the ensemble will increase to the accuracy. Additional complications 
can occur, if a rare canonical tunneling process (crossing of the 
barrier) takes actually place.

Let us compare with a multicanonical simulation. The multicanonical 
method flattens the barrier, whereas the parallel tempering simulation 
allows to jump it. Using larger lattices the multicanonical method is 
well suited for calculations of the interface tensions from energy 
histograms~\cite{BeNe92}. For parallel tempering this is not the case, 
because the sampling of each histogram is still canonical. This can
be overcome by a Gaussian variant of simulated tempering~\cite{NeHa06}.

\begin{figure} \centering
\includegraphics[height=4cm]{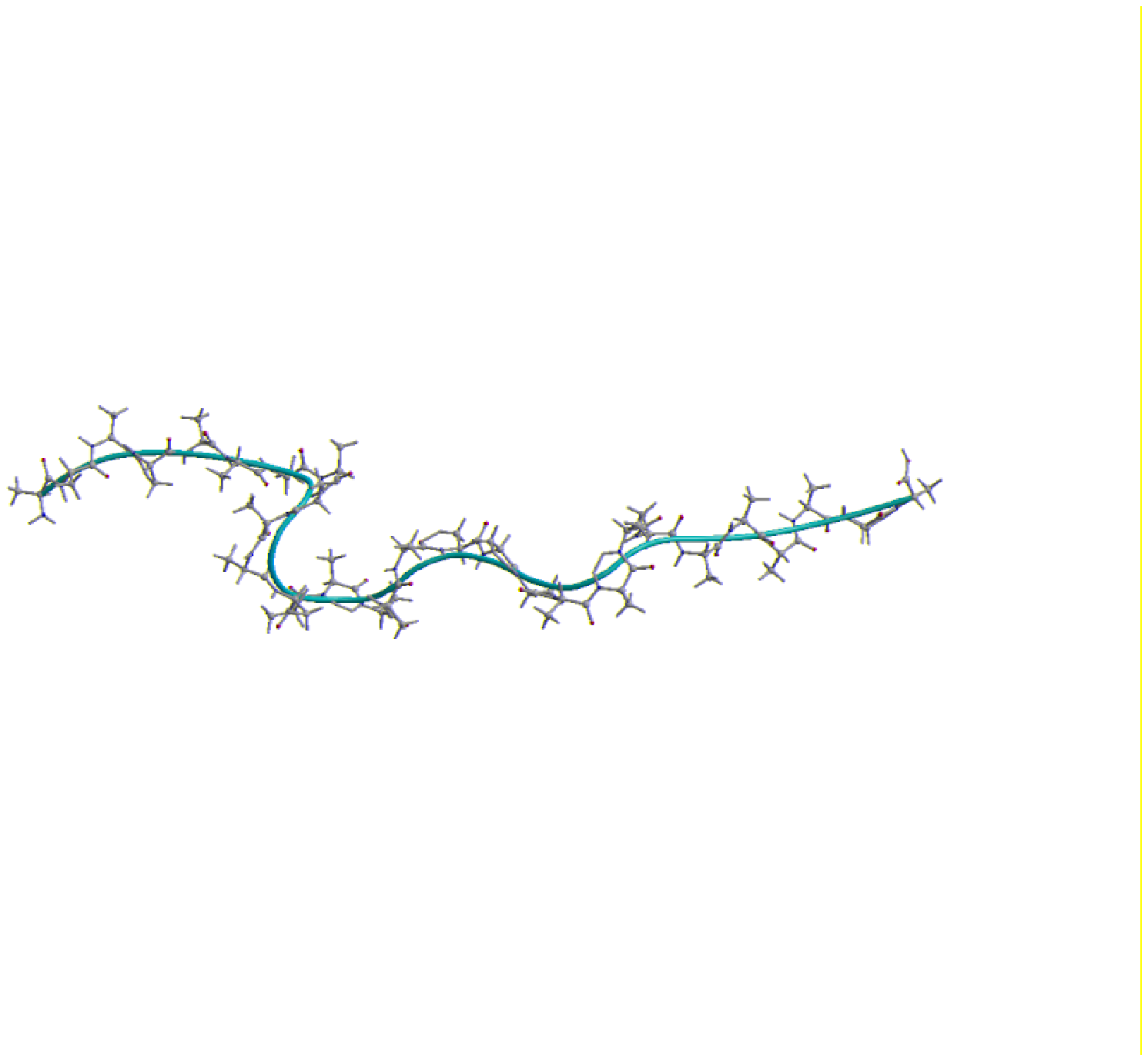}
\includegraphics[height=4cm]{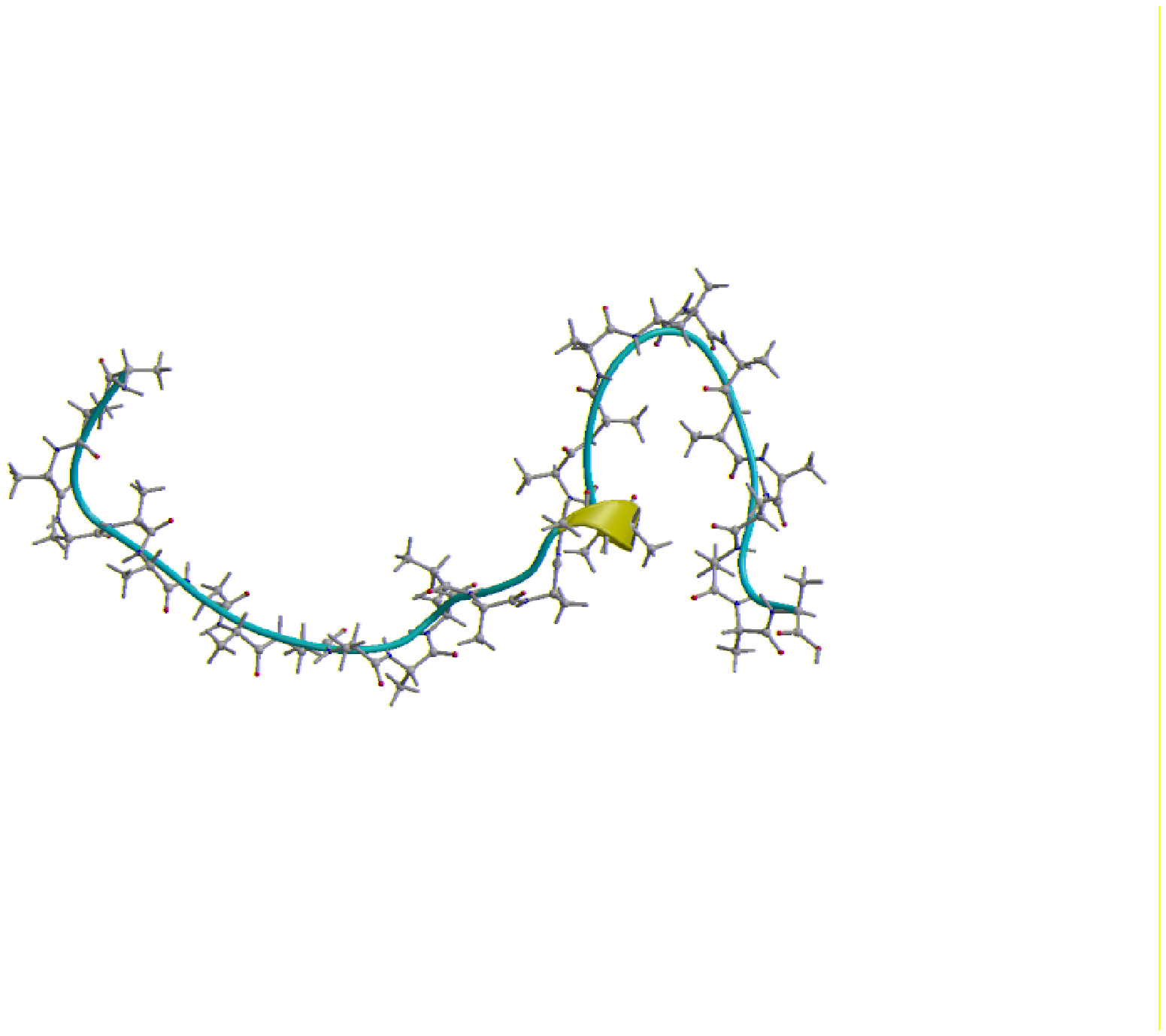}
\includegraphics[height=4cm]{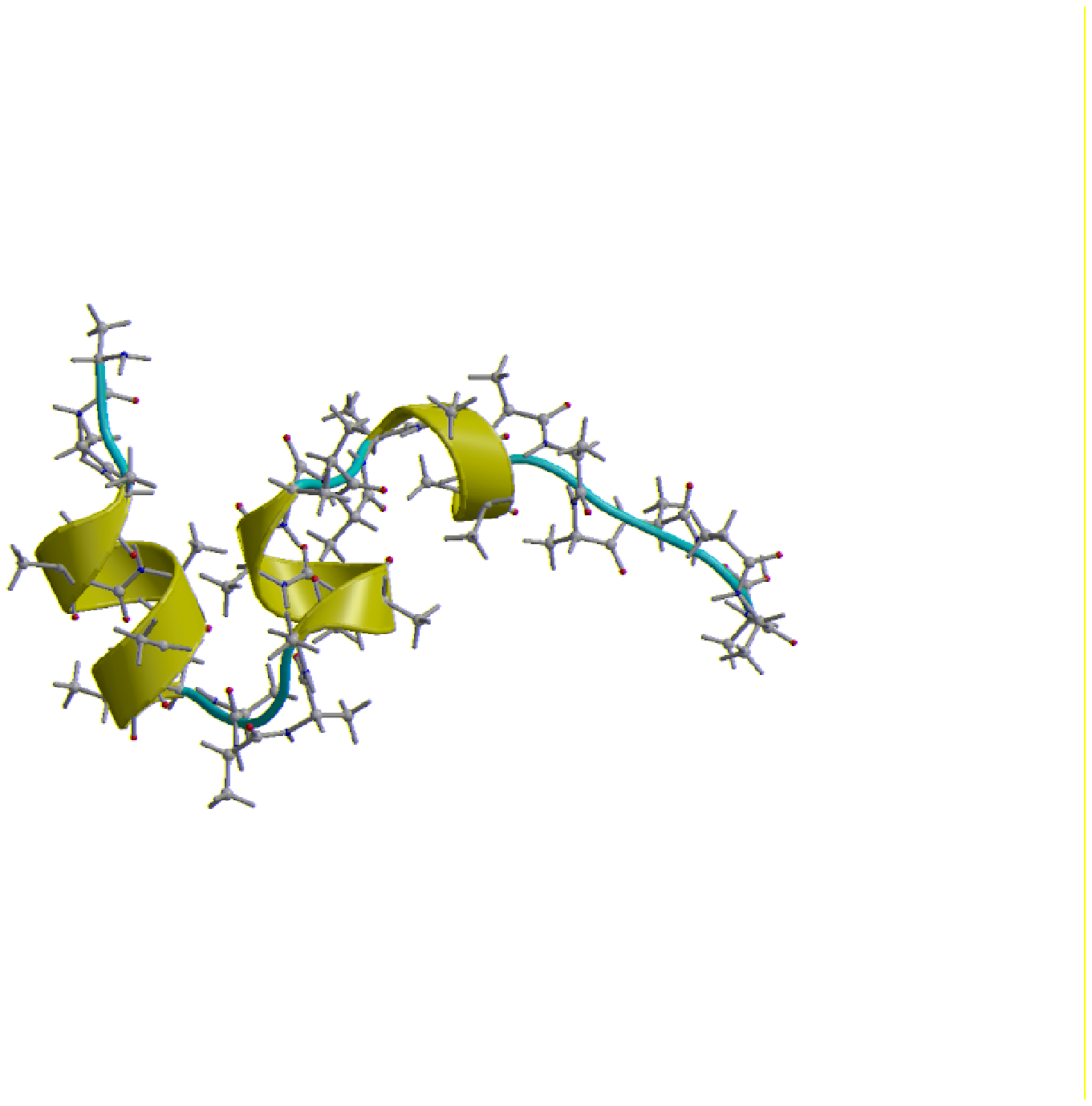}
\includegraphics[height=4cm]{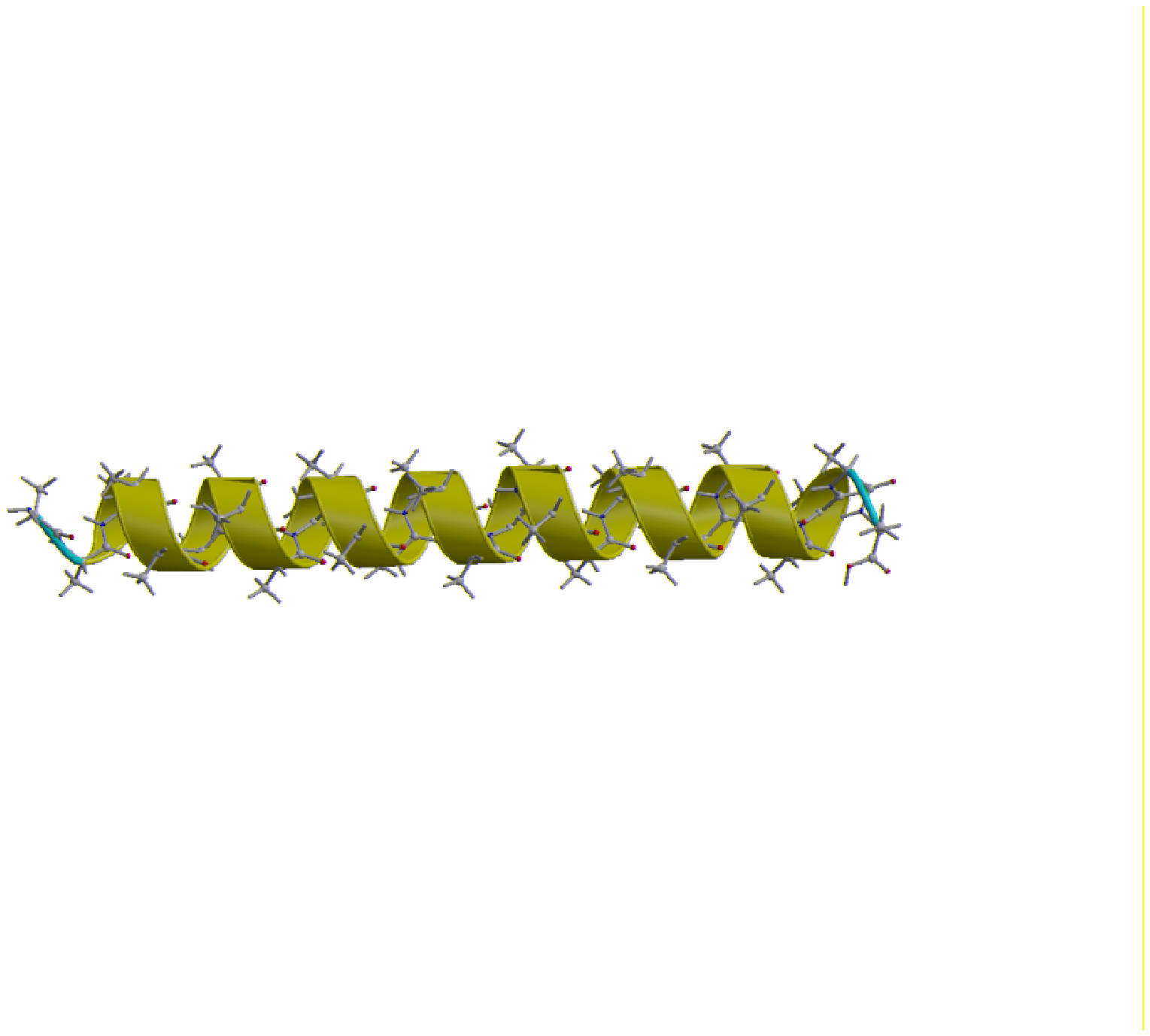}
\caption{Example configurations from a multicanonical simulation
of poly-alanine~\cite{HaOk99a} (courtesy Ulrich Hansmann and Yuko
Okamoto).} \label{fig_polyalanine}
\end{figure}

In complex systems with a rugged free energy landscape the barriers 
can no longer be explicitly controlled. Nevertheless it has turned out 
that switching to the discussed ensembles can greatly enhance the MCMC 
efficiency. A variety of biologically relevant applications are reviewed 
in Ref.~\cite{HaOk99,MiSu01} and in some of the lectures of this volume. 
Here we confine ourselves to showing a particularly nice example in 
Fig.~\ref{fig_polyalanine}: The folding of poly-alanine into its 
$\alpha$-helix coil~\cite{HaOk99a}. No \textit{a-priori} information 
about the groundstate conformation is used in these kind of simulations. 
Whereas in a canonical simulation one would in essence get stuck in a 
configuration like the second of the first row of this figure, the 
multicanonical simulation finds its way in and out of the helix 
structure. 

\begin{figure} \centering
\includegraphics[height=10cm]{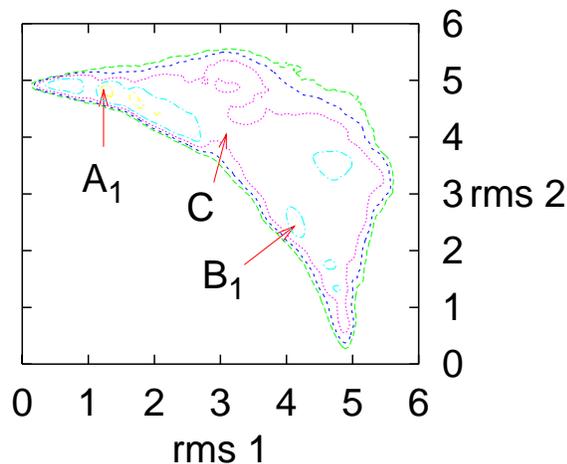}
\caption{Free energy landscape of Met-Enkephalin at $T=250$ K with 
respect to rms distances (\AA) from two reference configurations
\cite{BeNo03}.  \label{fig_F_rms} }
\end{figure}

Variations of the basic idea introduce weights in other variables than 
the energy (or even in several variables). For instance, weights in the
\index{magnetic field}magnetic field of spin systems were introduced 
quite some while ago~\cite{BeHa93}. For spin glasses weights in the 
Parisi overlap variable\index{Parisi overlap variable} are used
\cite{BeJa98}. The \index{overlap of configurations}overlap of the 
configuration at hand with two reference configurations allows one 
to determine the transition state\index{transition state} efficiently
\cite{BeNo03}. Fig.~\ref{fig_F_rms} shows a transition state located 
by this method in the free energy landscape of a simple peptide 
(Met-Enkephalin, which is discussed in the next section). In this 
figure contour lines are drawn every $2 k_B T$. The labels A$_1$ and 
B$_1$ indicate the positions for the local-minimum states that 
originate from the reference configuration~1 and the reference 
configuration~2, respectively. The label C stands for the saddle 
point that corresponds to the transition state. 

\subsection{Weight factor recursions}

For systems with rugged free energy landscapes FSS methods for 
estimating the \index{weights}weight factors are normally not 
available, because the systems are either of one particular size 
(e.g., biomolecules) or change too much when the system size is 
increased (e.g., spin glasses). In view of this one has to rely on 
ad-hoc per hand estimates or, more conveniently, on general purpose 
recursions. Designs for the latter were reported in a number of 
papers \cite{Be96,BeBi00,SuOk00,WaLa01a}. 

The recursions have in common that they intend to create weight factors 
which are inversely proportional to the spectral density of the system 
at hand (modifications to target other weights are straightforward). 
The Wang-Landau (WL) recursion\index{Wang-Landau recursion}
\cite{WaLa01a,WaLa01b} differs fundamentally from the other approaches 
by iterating the weight factor at energy $E$ multiplicatively instead 
of additively. At a first glance the WL approach is counter-intuitive, 
because the correct iteration of the weight factor after a sufficiently 
long simulation is obviously proportional to one over the number of 
histogram entries $H(E)$ and not $1/(f_{WL})^{H(E)}$ with $f_{WL}>1$. 
The advantage of the WL recursion is that it moves right away rapidly 
through the targeted energy range. Once this energy range is 
sufficiently covered, the WL method refines its iteration factor 
by $f_{WL}\to\sqrt{f_{WL}}$, so that it approaches~1. In this way 
the working approximations to the desired weight factors can be 
obtained on any finite energy range. In the following we give details 
for a variant~\cite{BBook} of the multicanonical 
recursion\index{multicanonical recursion} \cite{Be96} 
(the modifications rely on work with W. Janke) 
and the WL recursion~\cite{WaLa01a}. 

The multicanonical recursion uses the parameterization 
\cite{BeNe91,BeCe92} 
$$ w(a)\ =\ e^{-S(E_a)}\ =\ e^{-b(E_a)\, E_a + a(E_a)} $$
of the weights, where (for {$\epsilon$} smallest stepsize) 
\begin{eqnarray} \nonumber
  b(E) &=& 
 \left[ S(E+\epsilon) - S(E) \right] / \epsilon~~~{\rm and}\\
 \nonumber a(E-\epsilon) &=&  a(E) 
  + \left[ b(E-\epsilon)-b(E) \right]\, E\ . 
\end{eqnarray} 
After some algebra and other consideration \cite{BBook} the 
recursion reads
\begin{eqnarray} \nonumber
  b^{n+1}(E) & =&  b^n (E) + \hat{g}^n_0(E)\,
  [ \ln H^n(E+\epsilon)-\ln H^n(E)] / \epsilon\,, \\ \nonumber 
  \hat{g}^n_0 (E) &=& g^n_0(E)\, /\, [g^n(E) + \hat{g}^n_0 (E)]\,,
  \\ \nonumber 
  g^n_0 (E) &=& H^n (E+\epsilon)\, H^n (E)\, /\,
    [H^n (E+\epsilon) + H^n (E)]\,, \\ \nonumber
  g^{n+1} (E) &=& g^n(E) + g^n_0(E),\ g^0(E)=0\, .
\end{eqnarray}
For continuous systems like biomolecules some modification are 
required, see~\cite{YaCe00}.

For the WL recursion updates are performed with estimators $\rho(E)$ 
of the density of states
$$p(E_1\to E_2) = \min\left[ {\rho(E_1)\over \rho(E_2)}, 1\right]\ .$$
Each time an energy level is visited, they update the estimator
$$\rho(E) \to \rho(E)\,f_{WL} $$
where, initially, $\rho(E)=1$ and $f_{WL}=f_0=e^1$. Once the
desired energy range is covered, the factor $f_{WL}$ is refined to
$$f_1=\sqrt{f},\ f_{n+1}=\sqrt{f_{n+1}} $$
until some small value like $f_{WL}=e^{-8}=1.00000001$ is reached. 
For $f_{WL}$ very close to one the difference to a simulation with 
fixed weights becomes negligible, so that one may just keep on 
iterating $f_{WL}\to\sqrt{f_{WL}}$. However, it appears that such 
a continued iteration is far less efficient than switching to a 
simulation with fixed weights as soon as a working estimate is found.
Surprisingly there appears to be only one comparative study \cite{Ok03}
of the different recursions, which finds that overall performances are 
similar.

\section{Biased Markov Chain Monte Carlo}\label{sec_BMC}

Consider a random variable $y$ which is sampled with a probability
density function (PDF) $P(y)$ on an interval $[y_1,y_2]$. The 
cumulative distribution function (CDF) is defined by
\begin{equation}\label{basicCDF}
    z=F(y)=\int_{y_1}^y P(y')dy'\,\,\,\,\,
    \mbox{and}\,\,\,\,\,P(y)=\frac{dF(y)}{dy}\,,
\end{equation} 
where $P(y)$ is properly normalized so that $F(\infty)=1$ holds.  

The HBA generates $y$ by converting a uniformly distributed 
random number $0\leq z<1$ into
\begin{equation}\label{FHBA}
   y=F^{\,-1}(z)\, .
\end{equation} 
As the \textit{acceptance rate} is defined by the number of accepted
changes divided by the total number of proposed moves, the acceptance 
rate of the HBA is always 1 (a new value of $y$ is 
generated on \textit{every} step). In simulations the inversion of 
the CDF (\ref{basicCDF}) may be unacceptably slow or the CDF itself 
may not be \textit{a priori} known. Then one has to rely on other 
approaches.

In the conventional Metropolis scheme $y_{new}$ is generated uniformly 
in a range $[y_1,y_2]$ (we refer to this as \textbf{proposal}) 
\index{proposal}and accepted with probability 
\index{accept/reject step}(\textbf{accept/reject step})
\begin{equation} \label{pMet}
   p_{Met} = \min\left\{1,\frac{P(y_{new})}
   {P(y_{old})}\right\}.
\end{equation}
This process can have a low acceptance rate in the region of interest.
Possible remedies are to decrease the proposal range, which makes 
the moves small, or use of multi-hit Metropolis. Excluding CPU time 
considerations for the moment, measured by the integrated autocorrelation
time both remedies are normally less efficient than a HBA.

Hastings \cite{Ha70} identified proposal probabilities, which are
more general than those of the conventional Metropolis scheme, but
gave no guidance whether some probabilities may be preferable over 
others. If one does not propose $y_{new}$ uniformly anymore, the name 
\index{biased Metropolis Algorithm}\textbf{Biased Metropolis Algorithm 
(BMA)} is often used. Some biased Metropolis simulations can be found 
in the literature where the bias is introduced to match special 
situations \cite{BM1,BM2,BM3,BM4,BM5}. 
In the following we discuss a general biased Metropolis scheme 
\cite{Be03,BaBe05}, which aims at approximating heatbath 
probabilities. 

Let us discretize $y$ into $n$ bins as
\begin{equation}\label{ydiscr}
    y_1=y^0<y^1<y^2<...<y^n=y_2
\end{equation}
where the lengths of the bins are
\begin{equation}\label{deltay}
    \triangle y^j = y^j - y^{j-1},\,\,\,\,\,
    \mbox{with}\,\,\,\,\,j=1,...,n.
\end{equation}

A BMA can then be defined by the following steps:
\begin{itemize}
    \item Propose a new value $y_{new}$ by randomly picking
    a bin $j_{new}$ and then proposing $y_{new}$ uniformly in
    the given bin. Two uniform random numbers $r_1$, $r_2$ are needed:
    \begin{equation}\label{jnew}
        j_{new}=1+{\rm Int}[n\,r_1]\,\,\,\,\,\mbox{and}
        \,\,\,\,\,
        y_{new}=y^{j_{new}-1}+\triangle y^{j_{new}}\,r_2,
    \end{equation}
    where ${\rm Int}[n\,r_1]$ denotes rounding to the largest 
    integer $\le n\,r_1$.
    \item Locate the bin $j_{old}$ to which $y_{old}$ belongs: 
    \begin{equation}\label{jold}
        y^{j_{old}-1}\leq y_{old} \leq y^{j_{old}}.
    \end{equation}
    \item Accept $y_{new}$ with probability:
    \begin{equation} \label{pBMA}
        p_{BMA} = \min\left\{1,\frac{P(y_{new})}
        {P(y_{old})}\,
        \frac{\triangle y^{j_{new}}}{\triangle y^{j_{old}}}
        \right\}
    \end{equation}
\end{itemize}
$p_{BMA}$ in (\ref{pBMA}) differs from $p_{Met}$ in (\ref{pMet}) by
the \textit{bias} $\triangle y^{j_{new}}/\triangle y^{j_{old}}$. 
The scheme outlined in (\ref{jnew})-(\ref{pBMA}) satisfies balance 
or detailed balance in the same way as the original Metropolis 
algorithm, while the bias changes the acceptance rate.

So far the partitioning $y^j$ has not been introduced 
explicitly. A particular choice is:
\begin{equation}\label{jnF}
    \frac{j}{n}=F(y^j)\,\,\,\,\,\mbox{or}\,\,\,\,\,
    y^j=F^{-1}\left(\frac{j}{n}\right).
\end{equation}
This equation achieves equidistant partitioning on the CDF ordinate.
Let us pick a bin initially labeled $j$ and take the limit $n\to\infty$ 
so that this bin collapses into a point labeled $z$. Then this BMA 
approaches the HBA and the acceptance rate converges to~1: 
\begin{equation}\label{PoverP}
   \frac{P(y_{new})}{P(y_{old})}\,
   \frac{\triangle y^{j_{new}}}{\triangle y^{j_{old}}} 
   \to\frac{P(y_{new})} {P(y_{old})}\,
   \frac{1/P(y_{new})}{1/P(y_{old})}=1.
\end{equation} 
Therefore we call a BMA with the partitioning (\ref{jnF}) 
\textbf{Biased Metropolis-Heatbath Algorithm 
(BMA)}\index{biased Metropolis-Heatbath}.
Restricted to the updating of one dynamical variable the improvements
are similar to those seen in Fig.~\ref{fig_iathb}, where the gain is
a factor of five. Having in mind that that realistic simulations take 
usually months of computer time, such factors are highly welcome.
Extending the biased method to simultaneous updating of more than 
one variable, larger improvement factors can be anticipated. But 
extensions to more than two variables face technical difficulties.

\subsection{Application to a continuous model} 

Following Ref.~\cite{BaBe05} we illustrate the BMHA for a system with 
a continuous energy function: Pure lattice gauge theory calculations 
with the $U(1)$ gauge group\index{U(1) lattice gauge theory} for which 
the "matrices" are complex numbers on the unit circle, parameterized by 
an angle $\phi\in[0,2\pi)$. After defining the theory on the links of a 
four-dimensional lattice and going through some algebra, the PDF
\begin{equation}  \label{Pphi}
  P_\alpha(\phi) = N_\alpha\, e^{\alpha\,\cos (\phi)}
\end{equation} 
has to be sampled, where $\alpha$ is a parameter associated to the 
interaction of the link being updated with its environment. The 
corresponding CDF is
\begin{equation} \label{Fphi}
  F_\alpha(\phi) = N_\alpha \int_0^{\phi} d\phi'\,e^{\alpha\,
  \cos (\phi')} 
\end{equation} 
where $N_\alpha$ ensures the normalization $F_\alpha(2\pi)=1$. In the 
following we consider $U(1)$ gauge theory at a coupling close to the 
critical point for which one finds $0\leq\alpha\leq 6$. 

\begin{figure}[t]
\begin{center} 
\includegraphics[width=0.47\textwidth]{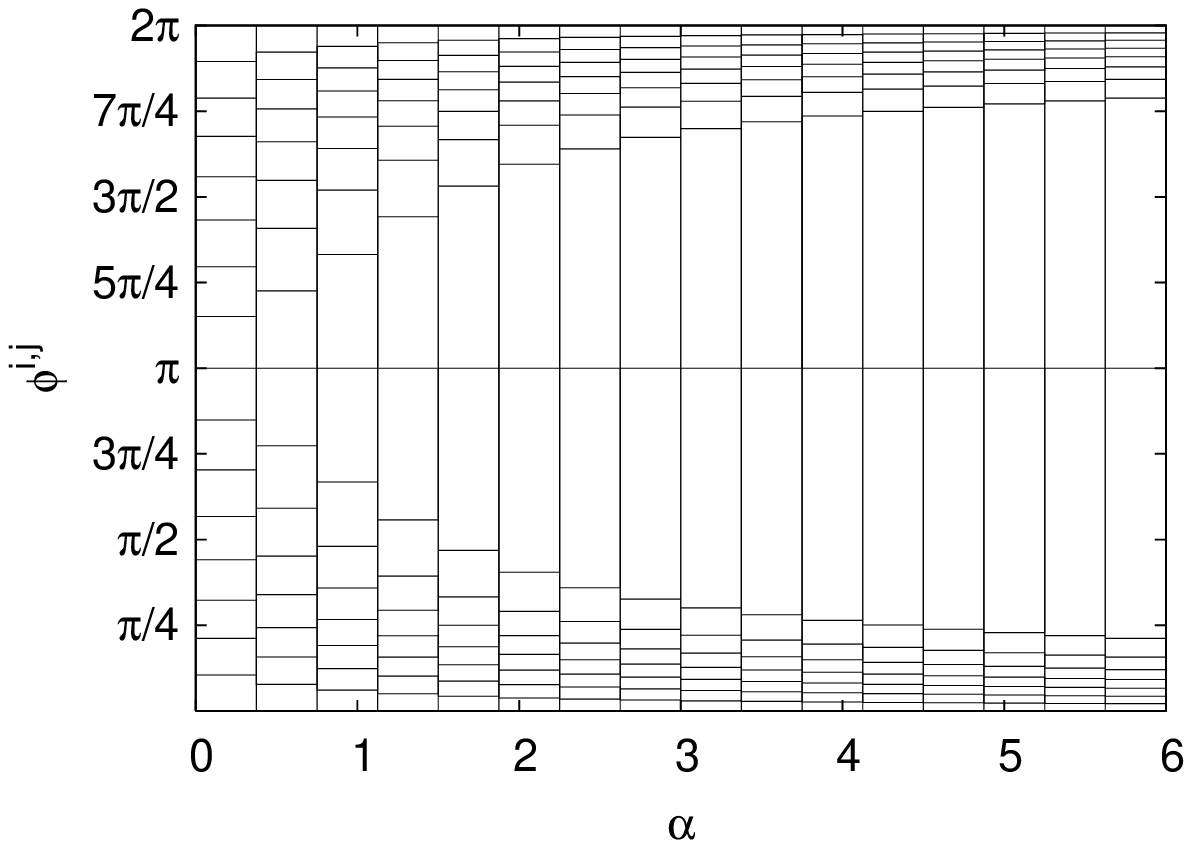}
\hfill
\includegraphics[width=0.47\textwidth]{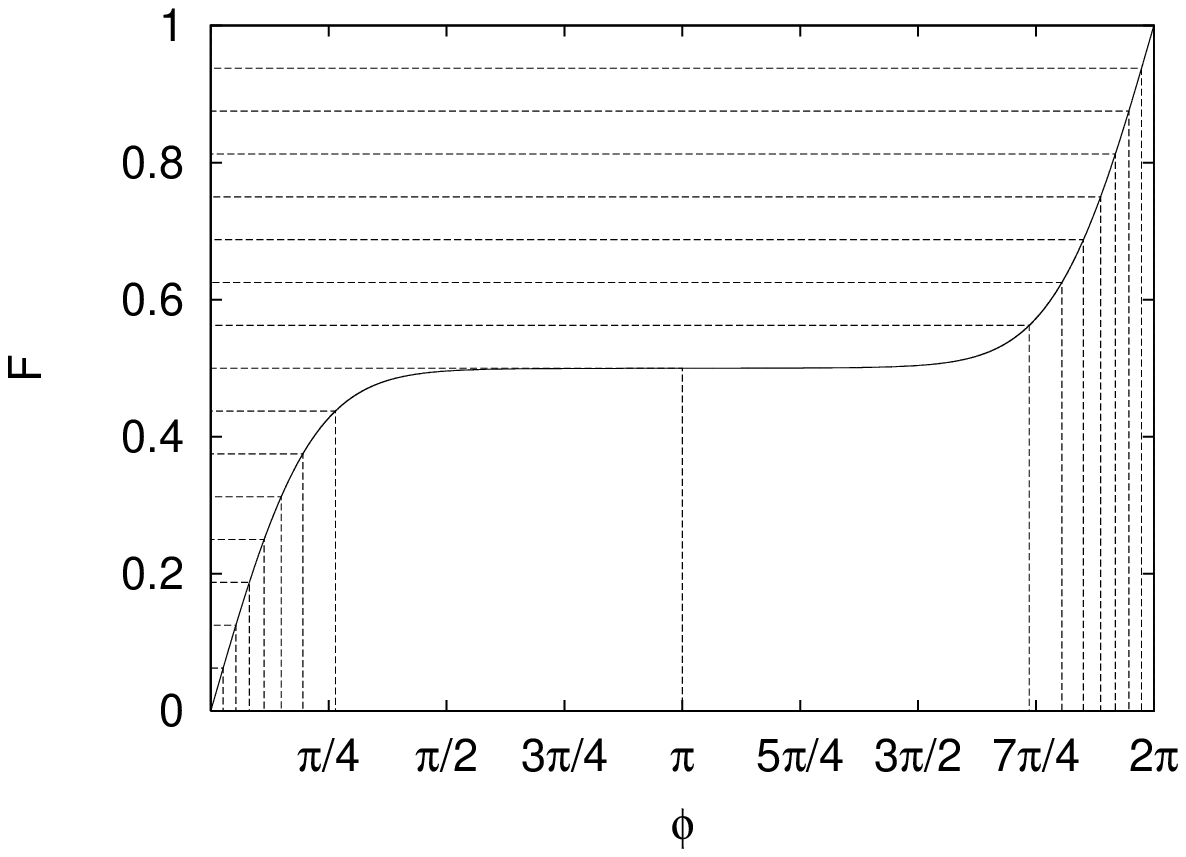}
\parbox[b]{0.47\textwidth}{\caption{ $m\times n$ partitioning of 
$\triangle \phi^{i,j}$ for $U(1)$ at the coupling constant value 
discussed in the text.}\label{fig_u1tab}}
\hfill
\parbox[b]{0.47\textwidth}{\caption{ Discretization of the CDF
$F_{\alpha^{11}}(\phi)$ for $U(1)$ corresponding to the 11th bin 
of Fig.~\ref{fig_u1tab}.}\label{fig_Fa0_u1}}
\end{center}
\end{figure}

Let us discretize the parameter $\alpha$ into $m=2^{n_1}=16$ ($n_1=4$) 
bins, choosing equidistant partitioning. In each $\alpha^i$ bin we 
discretize $\phi$ using the condition (\ref{jnF}) with $n=2^{n_2}=16$ 
($n_2=4$). Two two-dimensional arrays are needed: one for storing 
$\phi^{i,j}$ (levels themselves) and another for $\triangle \phi^{i,j} 
= \phi^{i,j} -\phi^{i,j-1}$ (distances between levels), see 
Fig.~\ref{fig_u1tab}. For a given $\alpha^i$ it is straightforward 
to apply BMA step (\ref{jnew}). E.g., for $\alpha=\alpha^{11}$, the 
cross section of the $F_{\alpha}(\phi)$ surface plane is then shown 
in Fig.~\ref{fig_Fa0_u1}. To determine the bin label $j_{old}$ which 
belongs to the (known) value $\phi_{0,old}$ (BMA step (\ref{jold})) 
one may use the $n_2$-step recursion $j~\to~j\,+\,2^{i_2}\ 
{\rm sign}\,(\,\phi-\phi^{i,j}\,)$, $i_2\to i_2-1$.
Once $j_{old}$ is known it gives the length of the bin: 
$\triangle \phi^{i,j_{old}}$ and the final accept/reject 
step (\ref{pBMA}) can be applied:
\begin{equation} \label{pBMA_U1}
   p_{BMA} 
   = \min\left\{1,\,\frac{\exp\left(\alpha\,\cos \phi_{0,new}\right)}
   {\exp\left(\alpha\,\cos \phi_{0,old}\right)}\,
   \frac{\triangle \phi_0^{i,j_{new}}}{\triangle \phi_0^{i,j_{old}}}\right\}.
\end{equation}
In this example the CDF is known. We have shown that sampling with the 
BMHA is essentially equivalent to using the HBA, but can be numerically 
faster, as is the case for $U(1)$. $SU(2)$ lattice gauge theory with 
the fundamental-adjoint action allows for substantial speed-ups by 
using a BMHA~\cite{BBH}. In the next section we show how a similar 
biasing procedure can be used when the CDF is not known (making a HBA 
impossible) and extend it to two variables.

\subsection{Rugged Metropolis, a biasing scheme for biophysics}
\label{sec_bio} 

We consider biomolecule models for which the energy $E$ is a function 
of a number of dynamical variables $v_i,\, i=1,\dots,n$. The 
fluctuations in the Gibbs canonical ensemble are described by a 
probability density function $\rho(v_1,\dots ,v_n; T)=const\,
\exp[-\beta\,E(v_1,\dots ,v_n)]$, where $T$ is the temperature, 
$\beta = 1 /(kT)$, and $E$ is the energy of the system. To be 
consistent with the notation of \cite{Be03} we now use 
$\rho(v_1,\dots ,v_n; T)$ instead of $P(y)$ introduced in previous 
one-variable example. Proposing a new variable (with the other 
variables fixed) from the PDF constitutes a HBA. However, an
implementation of a HBA is only possible when the CDF of the PDF
can be controlled. In particular this requires the normalization 
constant in front of the $\exp[-\beta\,E(v_1,\dots ,v_n)]$ Boltzmann 
factor, which is normally unknown. Then the following strategy can
provide a useful approximation.

For a range of temperatures
\begin{equation} \label{T_order}
  T_1\ >\ T_2\ >\ \dots\ >\ T_r\ >\ \dots\ >\ T_{f-1}\ >\ T_f
\end{equation} 
the simulation at the highest temperature, $T_1$, is performed 
with the usual Metropolis algorithm and the results are used
to construct an \textit{estimator} 
$$ {\overline\rho}(v_1,\dots,v_n;T_1)  $$
which is used to bias the simulation at $T_2$. Recursively, the 
estimated PDF
$$  {\overline\rho}(v_1,\dots,v_n;T_{r-1}) $$ 
is expected to be a useful approximation of $\rho(v_1,\dots,v_n;T_r)$. 
Formally this means that BMA acceptance step (\ref{pBMA}) at 
temperature $T_r$ is of the form
\begin{equation} \label{P0_acpt}
  P_{RM} = \min \left\{ 1, \frac{\exp\left(-\beta\,E'\right)
   }{ \exp\left(-\beta\,E\right)} \,\,
  \frac{{\overline\rho}(v_1,\dots,v_n;T_{r-1})}
  {{\overline\rho}(v'_1,\dots,v'_n;T_{r-1})}
   \right\}
\end{equation}
where $\beta=1/(kT)$. For this type of BMA, where the bias is 
constructed by using information from a higher temperature, the name
\index{rugged Metropolis}\textbf{Rugged Metropolis (RM)} was given 
in~\cite{Be03}.

Our test case in the following will is the small brain peptide 
\index{Met-Enkephalin}Met-Enkephalin (Tyr-Gly-Gly-Phe-Met) in vacuum, 
which features 24 dihedral angels as dynamical variables. Its global 
energy minimum was determined some time ago by Li and Scheraga 
\cite{LiSc87}. Ever since this molecule is often used as a simple 
laboratory for testing new computational methods. We rely on the 
all-atom energy function \index{ECEPP/2}ECEPP/2 (Empirical 
Conformational Energy Program for Peptides) \cite{SNS84} as 
implemented in the \index{SMMP}SMMP (Simple Molecular Mechanics 
for Proteins) \cite{Ei01} program package. Besides the $\phi,\,\psi$ 
angles, we keep also the $\omega$ angles unconstrained, which are 
usually restricted to $[\pi-\pi/9,\pi+\pi/9]$. This allows us to 
illustrate the RM idea for a particularly simple case.

To get things started, we need to construct an estimator 
${\overline\rho}(v_1,\dots,v_n;T_r)$ from the numerical data 
of the RM simulation at temperature $T_r$. This is neither simple 
nor straightforward, and approximations have to be used.

\subsubsection{The RM$_1$ approximation }

\begin{figure}[t]
\begin{center} 
\includegraphics[width=0.47\textwidth]{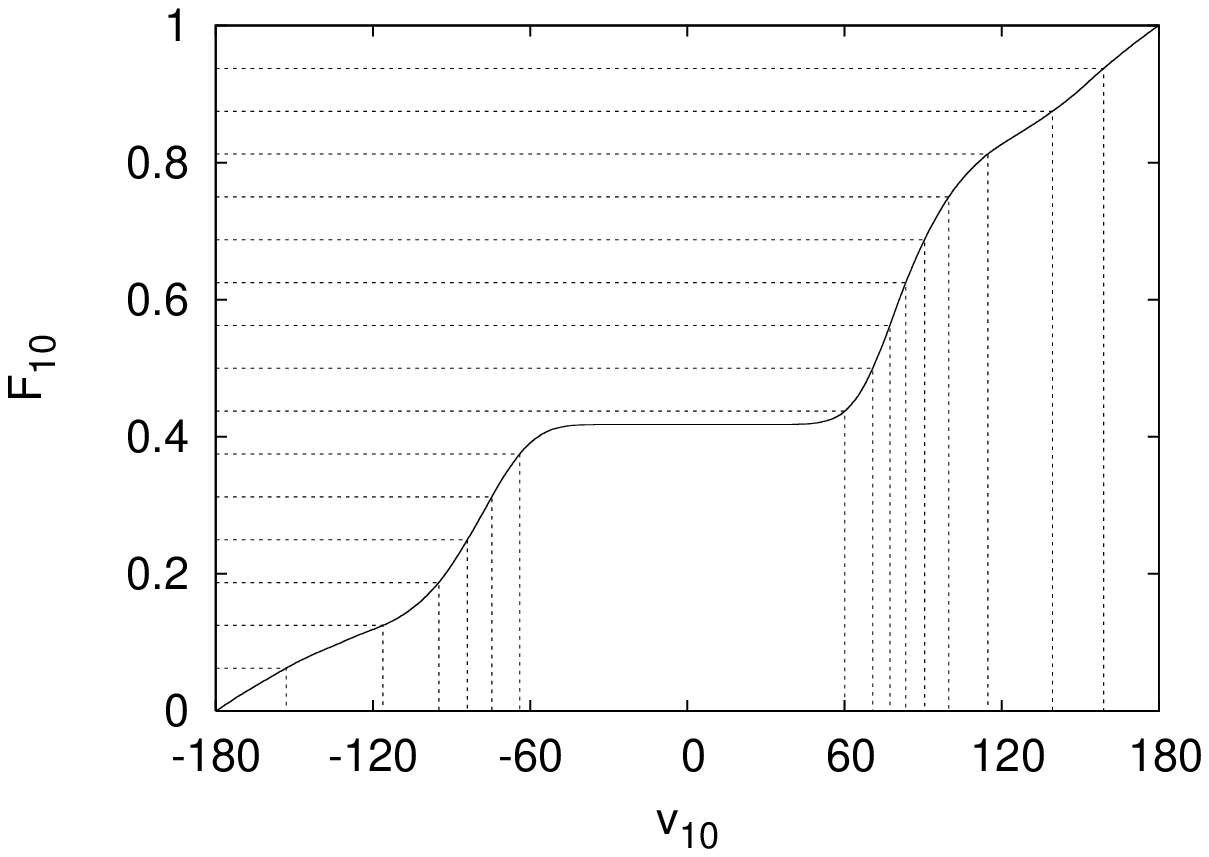}
\hfill
\includegraphics[width=0.47\textwidth]{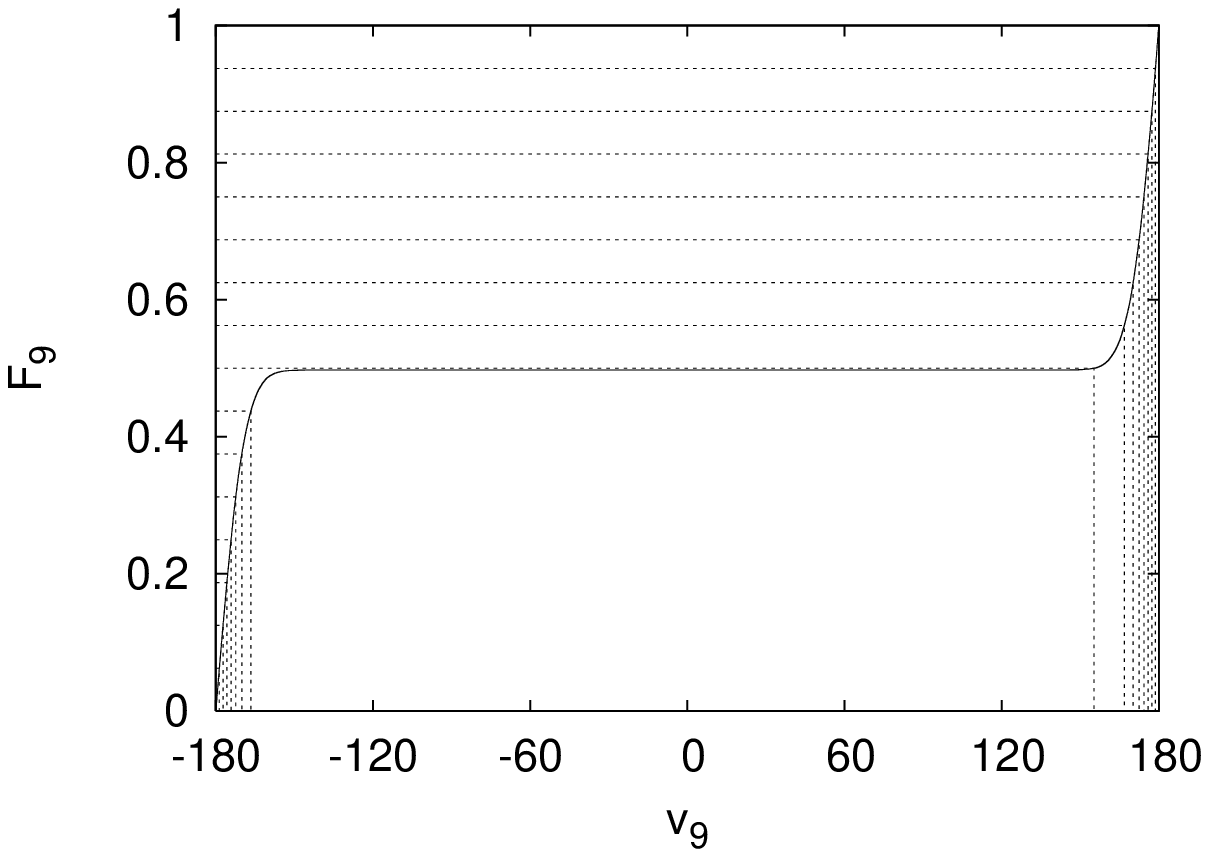}
\parbox[b]{0.47\textwidth}{\caption{Estimate of the cumulative 
distribution function for the Met-Enkephalin dihedral 
angle $v_{10}$ (Gly-3 $\phi$) at $400\,K$.}
\label{fig_Met_df10}}
\hfill
\parbox[b]{0.47\textwidth}{\caption{Estimate of the cumulative 
distribution function for the Met-Enkephalin dihedral 
angle $v_{9}$ (Gly-2 $\omega$) at $400\,$K.}
\label{fig_Met_df09}}
\end{center}
\end{figure}

In Ref.~\cite{Be03} the factorization
\begin{equation} \label{rho0_T0}
  {\overline\rho}(v_1,\dots,v_n;T_r) = \prod_{i=1}^n
  {\overline\rho}^1_i(v_i;T_r) 
\end{equation}
was investigated, where ${\overline\rho}^1_i(v_i;T_r)$ are 
estimators of reduced one-variable PDFs defined by
\begin{equation} \label{pd1}
 \rho^1_i(v_i;T) = \int_{-\pi}^{+\pi} \prod_{j\ne i} d\,v_j\,
 \rho(v_1,\dots ,v_n;T)\ .
\end{equation}
The resulting algorithm, called RM$_1$, constitutes the simplest RM 
scheme possible. The CDFs are defined by
\begin{equation} \label{df1}
  F_i(v)\ =\ \int_{-\pi}^v dv'\, \rho^1_i(v')\ .
\end{equation}
The estimate of $F_{10}$, the CDF for the dihedral angle Gly-3 
$\phi$ ($v_{10}$), from the simulations at our highest temperature, 
$T_1=400\,$K, is shown in Fig.~\ref{fig_Met_df10}. For our plots 
we use degrees, while we use radiant in our theoretical discussions 
and in the computer programs. Fig.~\ref{fig_Met_df10} is obtained by 
sorting all $n_{\rm dat}$ values of $v_{10}$ in our time series in 
ascending order and increasing the values of $F_{10}$ by $1/n_{\rm dat}$ 
whenever a measured value of $v_{10}$ is encountered. Using a heapsort 
\index{heapsort}\index{sorting}approach~\cite{BBook}, the sorting is 
done in $n_{\rm dat}\,\log_2(n_{\rm dat})$ steps.

Figure~\ref{fig_Met_df09} shows the CDF for $v_9$ (Gly-2 $\omega$) at 
$400\,K$, which is the angle of lowest 
acceptance rate in the conventional Metropolis updating. This 
distribution function corresponds to a histogram narrowly peaked 
around $\pm \pi$, which is explained by the specific electronic 
hybridization of the CO-N peptide bond. From the grid shown in 
Fig.~\ref{fig_Met_df09} it is seen that the RM$_1$ updating 
concentrates the proposal for this angle in the range slightly 
above $-\pi$ and slightly below $+\pi$. Thus the procedure has 
automatically a similar effect as the often used restriction to 
the range $[\pi-\pi/9,\pi+\pi/9]$, which is also the default 
implementation in SMMP.

After the empirical CDFs are constructed for each angle $v_i$, they 
are discretized using the condition (\ref{jnF}). Here we denote 
differences (\ref{deltay}) needed for the bias by
\begin{equation} \label{del_vij}
  \triangle v_{i,j}=v_{i,j}-v_{i,j-1}~~{\rm with}~~v_{i,0}=-\pi\ .
\end{equation} 
The RM$_1$ updating of each dihedral angle $v_i$ follows then the 
BMA procedure (\ref{jnew})-(\ref{pBMA}). The accept/reject step in
the $v_{i,j}$ notation is
\begin{equation} \label{pa_RM1}
    p_{RM_1} = \min\left\{ 1,
    \frac{ \exp (-\beta E') }{ \exp (-\beta E) }\,
    \frac{\triangle v_{i,j_{new}}}{\triangle v_{i,j_{old}}}
    \right\}\ .
\end{equation}

\subsubsection{The RM$_2$ approximation }

In Ref.~\cite{BeZh05} the RM$_1$ scheme of Eq.~(\ref{pa_RM1}) was
generalized to the \textit{simultaneous} updating of two dihedral 
angles. For $i_1\ne i_2$ reduced two-variable PDFs are defined by
\begin{equation} \label{pd2}
  \rho^2_{i_1,i_2}(v_{i_1},v_{i_2};T) = \int_{-\pi}^{+\pi} 
  \prod_{j\ne i_1,i_2} d\,v_j\, \rho(v_j,\dots ,v_n;T)\ .
\end{equation}
The one-variable CDFs $F_{i_1}$ and the 
discretization $v_{i_1,j},\,j=0,\dots,n$ are already given by 
Eqs.~(\ref{df1}) and~(\ref{del_vij}). We define conditional CDFs by
\begin{equation} \label{df2}
  F_{i_1,i_2;j}(v) = \int_{-\pi}^v dv_{i_2}
  \int_{v_{i_1,j-1}}^{v_{i_1,j}} dv_{i_1}\,
  \rho^2_{i_1,i_2}(v_{i_1},v_{i_2})
\end{equation}
for which the normalization $F_{i_1,i_2;j}(\pi)=1/n$ holds.
To extend the RM$_1$ updating to two variables we define for each 
integer $k=1,\dots,n$ the value $F_{i_1,i_2;j,k} =
k/n^2$. Next we define $v_{i_1,i_2;j,k}$ through 
$F_{i_1,i_2;j,k} = F_{i_1,i_2;j}(v_{i_1,i_2;j,k})$ and also the 
differences
\begin{equation} \label{del_vkj}
  \triangle v_{i_1,i_2;j,k} = v_{i_1,i_2;j,k} - v_{i_1,i_2;j,k-1}  
  ~~{\rm with}~~ v_{i_1,i_2;j,0} = -\pi\ .
\end{equation}
The RM$_2$ procedure for the simultaneous update of 
$(v_{i_1},v_{i_2})$ is then specified as follows:

\begin{itemize}
\item Propose a new value $v_{i_1,new}$ using two uniform random
      numbers $r_1$, $r_2$ (BMA step (\ref{jnew}) for the angle $i_1$):
      \begin{equation}\label{RM2_jnew}
        j_{new}=[n\,r_1]\,\,\,\,\,\mbox{and}
        \,\,\,\,\,
        v_{i_1,new}=v_{i_1,j_{new}-1}+\triangle v_{i_1,j_{new}}\,r_2.
      \end{equation}
\item Propose a new value $v_{i_2,new}$ using two uniform random
      numbers $r_3$, $r_4$ (BMA step (\ref{jnew}) for the angle $i_2$):
      \begin{equation}\label{RM2_knew}
        k_{new}=[n\,r_3]\,\,\,\,\,\mbox{and}
        \,\,\,\,\,
        v_{i_2,new}=v_{i_1,i_2;j_{new},k_{new}-1}+
        \triangle v_{i_1,i_2;j_{new},k_{new}}\,r_4.
      \end{equation}
\item Find the bin index $j_{old}$ for the present angle $v_{i_1,old}$ through
      $v_{i_1,j_{old}-1}\le v_{i_1,old}\le v_{i_1,j_{old}}$,
      just like for RM$_1$  updating (BMA step (\ref{jold}) for $v_{i_1}$).
\item Find the bin index $k_{old}$ for the present angle $v_{i_2,old}$ through
      $v_{i_1,i_2;j_{old},k_{old}-1}\le v_{i_2,old}
      \le v_{i_1,i_2;j_{old},k_{old}}$ (again step (\ref{jold})
      but for $v_{i_2}$).
\item Accept $(v_{i_1,new}$, $v_{i_2,new})$ with the probability 
      \begin{equation} \label{pa_RM2}
        p_{RM_2} = \min \left\{ 1 , 
        \frac{\exp (-\beta E')}{\exp (-\beta E)} \,
        \frac{\triangle v_{i_1,j_{new}}}{\triangle v_{i_1,j_{old}}} \,
        \frac{\triangle v_{i_1,i_2;j_{new},k_{new}}}
        {\triangle v_{i_1,i_2;j_{old},k_{old}}}
        \right\}\ .
      \end{equation}
\end{itemize}

As for RM$_1$, estimates of the conditional CDFs and the intervals 
$\triangle v_{i_1,i_2;j,k}$ are obtained from the conventional 
Metropolis simulation at 400$\,$K. In the following we focus on 
the pairs $(v_7,v_8)$, $(v_{10},v_{11})$ and $(v_{15},v_{16})$. These 
angles correspond to the largest integrated autocorrelation times of 
the RM$_1$ procedure and are expected to be strongly correlated with 
one another because they are pairs of dihedral angles around a 
$C_{\alpha}$ atom.

\begin{figure}[t]
\begin{center} 
\includegraphics[width=0.47\textwidth]{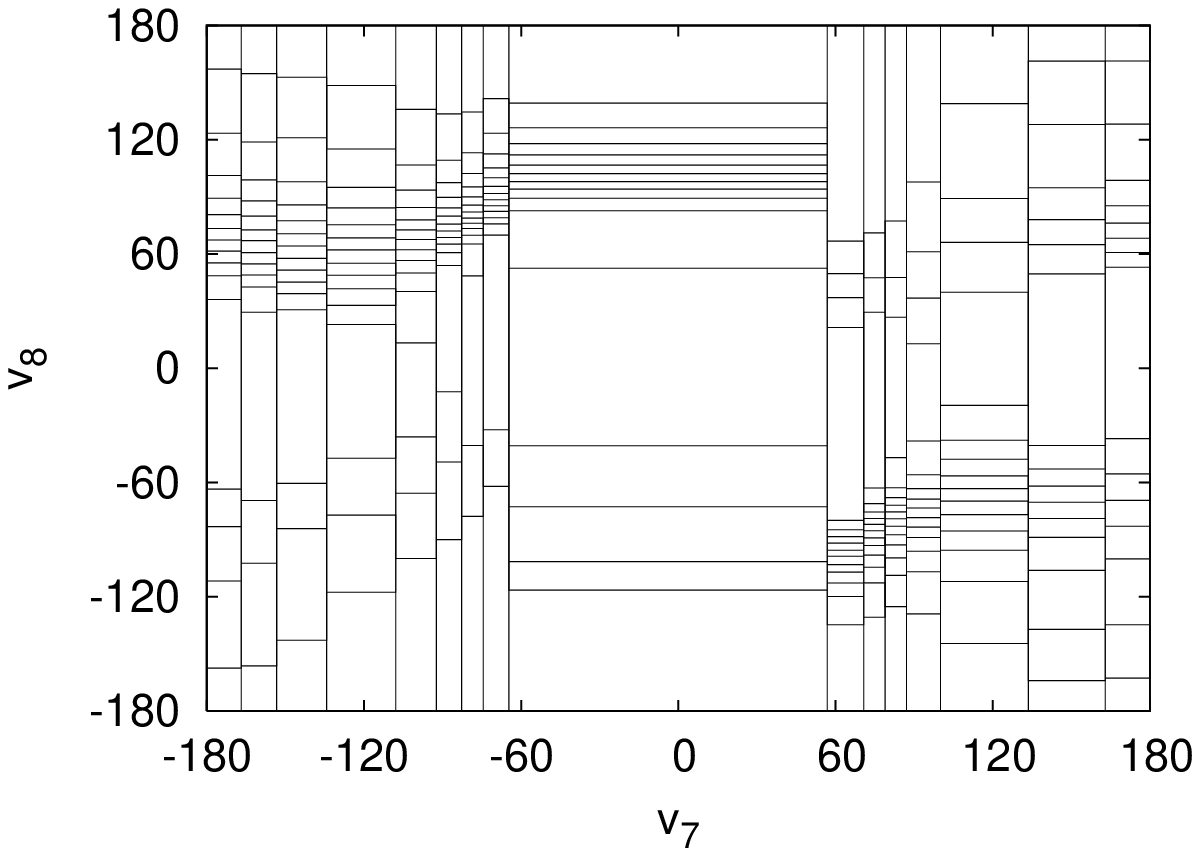}
\hfill
\includegraphics[width=0.47\textwidth]{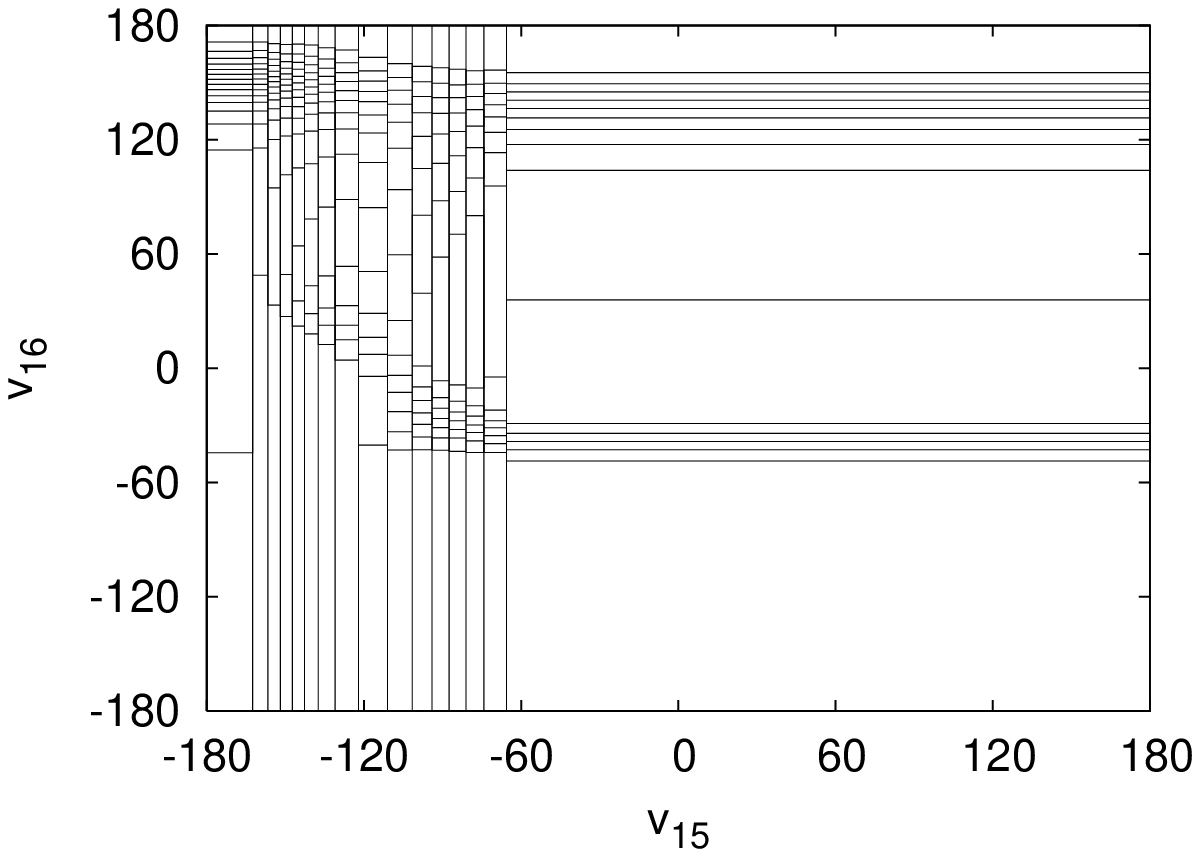}
\parbox[b]{0.47\textwidth}{\caption{Areas of equal probabilities
(sorting $v_7$ then $v_8$).}\label{fig_d07_d08}}
\hfill
\parbox[b]{0.47\textwidth}{\caption{Areas of equal probabilities
(sorting $v_{10}$ then $v_{11}$).}\label{fig_d15_d16}}
\end{center}
\end{figure}

The bias of the acceptance probability given in Eq.~(\ref{pa_RM2}) is 
governed by the areas 
$$ \triangle A_{i_1,i_2;j,k} = \triangle v_{i_1,j}\, 
   \triangle v_{i_1,i_2;j,k}\ . $$
For $i_1=7$ and $i_2=8$ our 400$\,$K estimates of these areas are
depicted in Fig.~\ref{fig_d07_d08}. For the RM$_2$ procedure these
areas take the role which the intervals on the abscissa of 
Fig.~\ref{fig_Met_df10} play for RM$_1$ updating. The small and 
the large areas are proposed with equal probabilities, so the 
a-priori probability for our two angles is high in a small area
and low in a large area. Areas of high probability correspond to 
allowed regions in the Ramachandran map of a Gly residue~\cite{Sch79}. 
In Fig.~\ref{fig_d07_d08} the largest area is 503.4 times the smallest 
area. Note that the order in which the angles are sorted introduces 
some differences. Fig.~\ref{fig_d15_d16} gives a plot for 
$(v_{15},v_{16})$ pairs in which the angle with the smaller 
subscript is sorted first. The ratio of the largest area over 
the smallest area is 2565.8. The large number is related to the 
fact that $(v_{15},v_{16})$ is the pair of $\phi,\,\psi$ angles 
around the $C_{\alpha}$ atom of Phe-4, for which positive $\phi$ 
values are disallowed~\cite{Sch79}.

Reductions in the integrated autocorrelations times of the angles vary 
and are again similar to those ovserved in Fig.~\ref{fig_iathb} when 
moving from the ordinary Metropolis to the HBA.

\section{Outlook and Conclusions}\label{sec_final}

Spin systems, lattice gauge theory and biophysics models are certainly 
far apart in their scientific objectives. Nevertheless quite similar 
computational techniques allow for efficient MCMC simulations in either
field. Cross-fertilization may go in both directions. For instance, 
generalized ensemble techniques propagated from lattice gauge theory 
\cite{BeNe91} over statistical physics \cite{BeCe92} into biophysics 
\cite{HaOk93}, while it appears that biased Metropolis techniques 
\cite{BM1,BM2,BM3,BM4,Be03} propagate in the opposite 
direction~\cite{BaBe05}. 

It is rather straightforward to combine the biased techniques of 
section~\ref{sec_BMC} with generalized ensembles, but work in this 
direction has just begun. In Ref.~\cite{Be03} and~\cite{BH04} 
combinations with parallel tempering have been studied and the
improvements were approximately multiplicative. An implementation
into the multicanonical ensemble has recently been worked 
out~\cite{BaBe06} and applied in a study of the deconfining 
phase transition of $U(1)$ lattice gauge theory. Extension of
the rugged MC approach to MD are also possible~\cite{WeBe06}.
As this leads into ongoing research, it is a good point to conclude
these lecture notes at this point.


%
%

\printindex
\end{document}